\documentclass[12pt,draftcls,onecolumn]{IEEEtran}

\usepackage{amsfonts}
\usepackage{cite}
\usepackage{graphicx}
\usepackage{amssymb}
\usepackage{subfigure}
\usepackage{array}
\usepackage{color}
\usepackage{amsmath}
\usepackage{algorithmic}
\usepackage{algorithm}
\usepackage{cases}
\usepackage{bm}
\usepackage{mathrsfs}
\usepackage{epsfig}
\usepackage{psfrag}
\usepackage{url}

\usepackage{etoolbox}
\makeatletter
\patchcmd{\@makecaption}
  {\scshape}
  {}
  {}
  {}
\makeatletter
\patchcmd{\@makecaption}
  {\\}
  {.\ }
  {}
  {}
\makeatother

\newcommand{\dis}{\stackrel{d}{\sim}}
\newcommand{\eqla}{\stackrel{(a)}{=}}
\newcommand{\eqlb}{\stackrel{(b)}{=}}
\newcommand{\eqlc}{\stackrel{(c)}{=}}
\newcommand{\eqld}{\stackrel{(d)}{=}}
\newcommand{\eqle}{\stackrel{(e)}{=}}
\newcommand{\eqlf}{\stackrel{(f)}{=}}

\newtheorem{Thm}{Theorem}
\newtheorem{Lem}{Lemma}
\newtheorem{Cor}{Corollary}

\newtheorem{Prob}{Problem}

\IEEEoverridecommandlockouts

\begin{document}

\title{Enhancing Performance of Random Caching in Large-Scale Wireless Networks with Multiple Receive Antennas}

\author{
\IEEEauthorblockN{Dongdong Jiang, {\em Student Member, IEEE} and Ying Cui, {\em Member, IEEE}} 
\thanks{D. Jiang and Y. Cui are with the Department of  Electronic Engineering, Shanghai Jiao Tong University, China. This paper will be presented in part at IEEE ICC 2018.}
}





\maketitle

\begin{abstract}
To improve signal-to-interference ratio (SIR) and make better use of  file diversity provided by random caching, we consider two types of linear receivers, i.e., maximal ratio combining (MRC) receiver and partial zero forcing (PZF) receiver, at users in a large-scale cache-enabled single-input multi-output (SIMO) network.
First, for each receiver, by utilizing tools from stochastic geometry, we derive a tractable expression and a tight upper bound for the successful transmission probability (STP).
In the case of the MRC receiver, we also derive a closed-form expression for the asymptotic outage probability  in the low SIR threshold regime.
Then, for each receiver, we maximize the STP.
In the case of the MRC receiver, we consider the maximization of the tight upper bound on the STP by optimizing the caching distribution, which is a non-convex problem. We obtain a stationary point, by solving an equivalent difference of convex (DC) programming problem using concave-convex procedure (CCCP). We also obtain a closed-form asymptotically optimal solution in the low SIR threshold regime.
In the case of the PZF receiver, we consider the maximization of the tight upper bound on the STP by optimizing the caching distribution and the degrees of freedom (DoF) allocation (for boosting the signal power), which is a mixed discrete-continuous problem.
Based on structural properties, we obtain a low-complexity near optimal solution by using an alternating optimization approach.
The analysis and optimization results reveal the impact of antenna resource at users on random caching.
Finally, by numerical results, we show that the random caching design with the PZF receiver achieves significant performance gains over the random caching design with the MRC receiver and some baseline caching designs.
\end{abstract}


\section{Introduction}

The rapid proliferation of smart mobile devices has triggered an unprecedented growth of the global mobile data traffic.
Motivated by the fact that a large portion of mobile data traffic is generated by many duplicate downloads of a few popular files,
recently, caching popular files at the wireless edge, namely caching helpers (base stations and access points) has been proposed as a promising approach for reducing delay and backhaul load \cite{Bustag16magazine,Shanmugam13,cachingmimoLiu15}. 
When the coverage regions of different helpers overlap, a user can fetch the desired file from multiple adjacent helpers, and hence the performance can be increased by caching different files among helpers, i.e., providing file diversity \cite{QuekTWC16}.
In~\cite{EURASIP15Debbah,Hassan15,Bharath16TCOM,Cui16TWC,Cui16Hetnet}, the authors consider caching in large-scale networks modeled using stochastic geometry. 
Specifically, in~\cite{EURASIP15Debbah}, 
the authors consider caching the most popular files at each helper, which does not provide file diversity.
In~\cite{Hassan15}, the authors consider random caching with uniform distribution at each helper. 
In~\cite{Bharath16TCOM}, the authors consider random caching with files being stored at each helper in an i.i.d. manner according to the file popularity.
In~\cite{Cui16TWC} and \cite{Cui16Hetnet}, the authors consider random joint caching and multicasting on the basis of file combinations consisting of different files, and analyze and optimize the joint design.
Note that the random caching schemes in \cite{Hassan15,Bharath16TCOM,Cui16TWC,Cui16Hetnet} can provide file diversity.
However, in \cite{Hassan15,Bharath16TCOM,Cui16TWC,Cui16Hetnet}, a user may associate with a relatively farther helper when nearer helpers do not cache the requested file. In this case, the signal is usually weak compared with the interference, and the user may not successfully receive the requested file and benefit from file diversity offered by random caching.

To increase signal-to-interference ratio (SIR) under random caching, \cite{Wen17arXiv,ChonMobiHoc15,Liu17arXiv,Jifang17} study more sophisticated transmission schemes in large-scale networks. In particular, to increase receive signal power under random caching, \cite{Wen17arXiv,ChonMobiHoc15,Liu17arXiv} consider cooperative transmission schemes, where multiple nearest helpers \cite{Wen17arXiv,ChonMobiHoc15} or all helpers \cite{Liu17arXiv} storing the requested file of a user transmit it to the user, using non-coherent joint transmission~\cite{Wen17arXiv,Liu17arXiv} and coherent joint transmission~\cite{ChonMobiHoc15}.
The successful transmission probability (STP) \cite{Wen17arXiv,Liu17arXiv} and the effective transmission rate \cite{ChonMobiHoc15} are analyzed and optimized.
To reduce interference under random caching, in~\cite{Jifang17}, the authors propose a periodic discontinuous transmission scheme, where each helper is on once every $L$ slots and serves all file requests received during the latest $L$ slots via multicast, reducing the interference to $1/L$ of that with all helpers being on. The STP is analyzed and optimized. 
Note that the cooperative transmission schemes in \cite{Wen17arXiv,ChonMobiHoc15,Liu17arXiv} are applicable mainly in lightly loaded networks, where multiple helpers can jointly serve a single user;
the periodic discontinuous transmission scheme in~\cite{Jifang17} increases the STP at the cost of delay increase.

Note that \cite{Wen17arXiv,ChonMobiHoc15,Liu17arXiv,Jifang17} focus on improving SIR for SISO transmission with single-antenna transmitters and receivers. 
Multi-antenna communication techniques provide another promising approach to increase SIR. 
In \cite{Lee15TWC} and \cite{CuiTWC16IN}, the authors consider multi-input single-output (MISO) transmission, where each multi-antenna transmitter adopts a coordinated beamforming scheme to nullify inter-cell interference and increase receive signal power at single-antenna receivers,  and analyze the coverage probability in large-scale wireless networks.
In addition, in~\cite{LeeTIT16,JindalTCOM11,AliTWC10}, the authors consider single-input multi-output (SIMO) transmission, where maximal ratio combining (MRC) receiver~\cite{LeeTIT16}, partial zero-forcing (PZF) receiver~\cite{JindalTCOM11} and minimum mean-square error (MMSE) receiver~\cite{AliTWC10} are adopted at each multi-antenna receiver to increase the receive signal power or suppress the interference, 
and analyze  the scaling law of the spectrum efficiency~\cite{LeeTIT16}, the scaling law of the transmission capacity~\cite{JindalTCOM11} and the outage probability~\cite{AliTWC10} in large-scale wireless networks.
Note that \cite{Lee15TWC,CuiTWC16IN,LeeTIT16,JindalTCOM11,AliTWC10} do not consider caching. 
Recently, \cite{kuang2017random} and \cite{LiuTCOM17} focus on improving SIR under  random caching in large-scale MISO networks. Specifically, in \cite{kuang2017random}, the authors consider maximal ratio transmission (MRT) beamforming to improve receive signal power, and analyze and optimize the area spectrum efficiency. 
In~\cite{LiuTCOM17}, the authors consider zero-forcing beamforming (ZFBF) to simultaneously serve multiple single-antenna users without causing intra-cell interference, and analyze and optimize the STP and the area spectral efficiency.
Note that the MRT beamforming scheme in~\cite{kuang2017random} and the ZFBF scheme in~\cite{LiuTCOM17} require perfect CSI at transmitters. In addition, \cite{kuang2017random} does not consider interference management, and \cite{LiuTCOM17} studies interference management which may not be very efficient from the user's point of view. Finally,
in~\cite{kuang2017random}, 
the performance expression involves inverse of matrices, which is difficult to evaluate and provides little insight, and the adopted gradient projection method for solving the non-convex caching optimization problem has high computation complexity and possibly slow convergence;  
in \cite{LiuTCOM17}, the authors analyze and optimize the STP and the area spectrum efficiency under several approximations.

Note that \cite{kuang2017random} and \cite{LiuTCOM17} focus on revealing the impact of antenna resource at transmitters on improving SIR under random caching. It is not known whether antenna resource at receivers can achieve a similar or even more important role in SIR improvement under random caching.
In this paper, we would like to investigate how multiple receive antennas at users can help improve SIR for content delivery when single-antenna helpers employ random caching in a large-scale cache-enabled SIMO network. 
Specifically, we consider $M$-antenna users and study two types of linear receivers for two cases of CSI at users. First, we consider the direct CSI case where each user has knowledge of the CSI of the direct link between its serving helper and itself, and perform MRC at each user to increase the receive signal power. Next, we consider the local CSI case where each user has knowledge of the CSI of the links between its nearby (at most $M-1$ nearest) interfering helpers and itself as well as the direct link, and perform PZF at each user to simultaneously increase the receive signal power and suppress the interference. 
Note that the two receive schemes only require CSI at receivers and do not require CSI at transmitters. In addition, the PZF receiver can suppress interference from nearby interferers more efficiently from the user's point of view.
Our main contributions are summarized bellow.
\begin{itemize}

\item First, we analyze the STP. In the case of the MRC receiver, by utilizing tools from stochastic geometry and inequalities for the incomplete gamma function, we derive a tractable expression and closed-form upper and lower bounds for the STP. 
    The two bounds are tight when $M=1$, and the upper bound is shown to be a good approximation for the STP at all $M$. We also derive a closed-form expression for the asymptotic outage probability in the low SIR threshold regime, utilizing series expansion of some special functions.
    In the case of the PZF receiver, by utilizing tools from stochastic geometry and relative locations of the serving helper and interferers, we derive a tractable expression and a simpler upper bound for the STP.
    The upper bound is shown to be a good approximation for the STP at all $M$.
    The analysis results reveal that the STPs with the MRC and PZF receivers both increase with $M$.

\item  Next, we maximize the STP. 
    In the case of the MRC receiver, we consider the maximization of the simpler upper bound on the STP by optimizing the caching distribution, which is a non-convex problem. 
    By exploring structural properties of the problem, 
    we successfully transform the original non-convex problem into a difference of convex (DC) programming problem, and obtain a stationary point of the original problem, using concave-convex procedure (CCCP).  We also obtain a closed-form asymptotically optimal solution in the low SIR threshold regime.
    In the case of the PZF receiver, we consider the maximization of the simpler upper bound on the STP by optimizing the caching distribution and the degrees of freedom (DoF) allocation, which is a mixed discrete-continuous problem. By exploring structural properties of the problem, we obtain a low complexity near optimal solution by an alternating optimization approach. 
    The optimization results indicate that files of higher popularity get more storage resources, and
    the optimized caching distributions for the MRC and PZF receivers both become more flat when $M$ is larger. 

\item Finally, we show that the proposed random caching design with the PZF receiver achieves significant performance gains over the proposed random caching design with the MRC receiver and some baseline caching schemes, using numerical results.
\end{itemize}

\section{System Model and Performance Metric}


\subsection{Network Model}\label{subsec:net_model}

\begin{figure}[t]
\begin{center}
 \includegraphics[width=4.5cm]{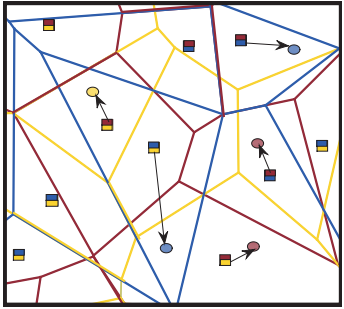}
  \end{center}
  \vspace{-6mm}
  \caption{\small{System model ($N=3$, $C=2$).
  There are three files in the network, represented by the red, yellow and blue colors, respectively. Each circle represents a user, the color of which indicates the file requested by the user. Each square represents a caching helper, the color of which indicates the two different files stored at the helper~\cite{Cui16TWC}. 
  }}
  \vspace{-8mm}
\label{fig:system}
\end{figure}

We consider a large-scale cache-enabled network,\footnote{The system model is similar to the one we considered in \cite{Cui16TWC}, except that here we consider multi-antenna users. Here, we briefly illustrate the system model for completeness.} as shown in Fig.~\ref{fig:system}. The locations of caching helpers are spatially distributed as a two-dimensional homogeneous Poisson point process (PPP) $\Phi_{h}$ with density $\lambda_{h}$. The locations of users are distributed as an independent two-dimensional homogeneous PPP with density $\lambda_u$. According to Slivnyak's theorem \cite{haenggi2012stochastic}, we focus on a typical user $u_0$, which we assume without loss of generality (w.l.o.g.) to be located at the origin. The helpers are labeled in ascending order of distance from $u_0$. Let $d_i$ denote the distance between helper  $i\in \Phi_h$ and $u_0$. Thus, we have $d_1\leq d_2\leq \cdots$. We consider the downlink transmission.
Each helper has one transmit antenna with transmission power $P$. All helpers transmit over the same frequency band. Each user has $M$ receive antennas. That is, we focus on SIMO transmission.\footnote{Note that the analysis and optimization results in this paper can be extended to MIMO transmission with open-loop spatial multiplexing, by treating one multi-antenna helper sending multiple data streams as multiple co-located virtual single-antenna helpers, each sending one data stream, as illustrated in \cite{Lin15TWC}.} 
Due to path loss,  transmitted signals with distance $d$ are attenuated by a factor $d^{-\alpha}$, where $\alpha>2$ is the path loss exponent. 
Let $\mathbf h_{i,0}\in\mathbb C^{M\times 1}$ denote the small-scale fading vector between helper $i\in\Phi_h$ and $u_0$. We assume that all entries of $\mathbf h_{i,0}$ are i.i.d. complex Gaussian random variables, each with zero mean and unit variance, i.e., $\mathcal {CN}(0, 1)$.

Let $\mathcal N\triangleq \{1,2,\cdots, N\}$ denote the set of $N$ files in the network. For ease of illustration, we assume that all files have the same size. The popularity distribution among $\mathcal N$  is assumed to be known apriori and is denoted by $\mathbf a\triangleq (a_n)_{n\in \mathcal N }$. That is, $u_0$ randomly requests one file, which is file $n\in \mathcal N$ with probability $a_n$, where $\sum_{n\in \mathcal N}a_n=1$.  In addition, w.l.o.g., we assume $a_{1}> a_{2}\ldots> a_{N}$.

\subsection{Random Caching Design}

The network consists of cache-enabled helpers. In particular, each helper is equipped with a cache of size $C\geq 1$ (in files) and can serve any files stored locally. Assume each helper cannot store all files in $\mathcal N$ due to the limited storage capacity, i.e., $C<N$. To provide spatial file diversity (which can improve performance of dense wireless networks), we adopt a random
caching design at helpers~\cite{Cui16TWC}. In particular, each helper stores $C$ different files out of all $N$ files in $\mathcal N$ with a certain probability. Let $T_n$ denote the probability of file $n$ being stored at each helper. Then, we have
\small{\begin{align}
&1\leq T_n \leq 1,\quad n\in\mathcal N,\label{eqn:cache_interval_constr}\\
&\sum_{n\in\mathcal N}T_n = C.\label{eqn:cache_size_constr}
\end{align}}\normalsize
Denote $\mathbf T\triangleq (T_n)_{n\in\mathcal N}$, which is termed as the caching distribution.
In this paper, we focus on serving cached files  to get first-order insights into the design of cache-enabled SIMO wireless networks. 

Each user requesting a file is associated with the nearest helper storing this file, referred to as its serving helper, as this helper offers the maximum long-term average receive power for this file. Suppose $u_0$ requests file $n$. Let $\ell_{0,n}\in\Phi_h$ denote the index of the serving helper of $u_0$. Note that the serving helper of $u_0$ may not be its geographically nearest helper, and the distance between $u_0$ and its serving helper is statistically determined by $T_n$. This association mechanism is referred to as the content-centric association~\cite{Cui16TWC}. Different from the traditional connection-based association~\cite{Lin15TWC}, this association jointly considers the physical layer and content-centric properties. 

\subsection{Receive Filters}

For analytical tractability, as in~\cite{QuekTWC16}, \cite{EURASIP15Debbah} and \cite{Lin15TWC}, we assume that all helpers are active for serving their own users (e.g., in the case of $\lambda_u\gg\lambda_h$).
Suppose $u_0$ requests file $n$. Then, the received signal of $u_0$, 
denoted as $\mathbf y_n\in\mathbb C^{M\times 1}$, is given by
\small{\begin{align}
\mathbf y_n = d_{\ell_{0,n}}^{-\frac{\alpha}{2}}\mathbf h_{\ell_{0,n},0}s_{\ell_{0,n}}+\sum_{i\in\Phi_h\setminus\{\ell_{0,n}\}}d_i^{-\frac{\alpha}{2}}\mathbf h_{i,0}s_i+\mathbf z_0,\notag
\end{align}}\normalsize
where $s_i$ is the transmit signal of helper $i$ with $\mathbb E[\|s_i\|^2]=P$ and $\mathbf z_0\in\mathbb C^{M\times 1}$ is complex Gaussian noise vector at $u_0$. As in~\cite{QuekTWC16}, in the following, we consider an interference-limited case (i.e., the noise power is negligible compared to the interference power) and ignore the noise.

Denote $\mathbf w_0\in\mathbb C^{M\times 1}$ as the receive filter adopted by $u_0$ for receiving the signal of its serving helper. That is,  $u_0$ detects signal $s_{\ell_{0,n}}$ based on $\mathbf w_{0,n}^H\mathbf y_n$, where $\mathbf w_{0,n}^H$ denotes the conjugate transpose of vector $\mathbf w_{0,n}$. The performance of the cache-enabled SIMO network depends on the choices of the receive filter. In the following, we consider two types of linear receivers for two different CSI cases, respectively.

First, we consider the case where each user has only knowledge of the CSI of its direct link, i.e., the link between its serving helper and itself, referred to as direct CSI.
Obtaining direct CSI 
can be done using a control channel with a reasonable amount of pilot signal overhead~\cite{LeeTIT16}. Given direct CSI only, we consider the MRC receiver, which is 
the optimal receiving strategy to maximize the signal power in SIMO transmission.
In particular, the MRC receive filter is given by $\mathbf w_{0,n}^{\rm mrc}=\frac{\mathbf h_{\ell_{0,n},0}}{\|\mathbf h_{\ell_{0,n},0}\|_2}$,
and the corresponding SIR of $u_0$ is given by
\small{\begin{align}
\text{SIR}_{n}^{\text{mrc}}=\frac{H_{\ell_{0,n},0}^{\text{mrc}}d_{\ell_{0,n}}^{-\alpha}}{\sum_{i\in \Phi_h \setminus\{\ell_{0,n}\}}H_{i,0}^{\text{mrc}}d_i^{-\alpha}},\label{eqn:SIR_MRC}
\end{align}}\normalsize
where 
$H_{\ell_{0,n},0}^{\text{mrc}} \triangleq  \|(\mathbf w_{0,n}^{\rm mrc})^H\mathbf h_{\ell_{0,n},0}\|_2^2 = \|\mathbf h_{\ell_{0,n},0}\|_2^2$ denotes the fading power of the direct link, and $H_{i,0}^{\text{mrc}} \triangleq \|(\mathbf w_{0,n}^{\rm mrc})^H\mathbf h_{i,0}\|_2^2$ denotes the fading power of the link between helper $i\in\Phi_{h}\setminus\{\ell_{0,n}\}$ and $u_0$. Here, $H_{\ell_{0,n},0}^\text{mrc}\dis \text{Gamma}(M,1)$ and $H_{i,0}^{\text{mrc}}\dis \text{Exp}(1)$ \cite{JindalTCOM11}. 
Note that MISO transmission with the MRT beamformer can achieve a SIR of the same distribution as the one in \eqref{eqn:SIR_MRC}, based on CSI at transmitters \cite{kuang2017random}.

Next, we consider the case where each user is able to learn the CSI of the links between its nearby (at most $M-1$ nearest) interfering helpers and itself as well as the direct link, referred to as local CSI. 
Given local CSI, we adopt a PZF receiver, which uses a subset of the $M$ DoF for boosting signal power and the remainder for interference cancellation. 
In particular, at $u_0$, $K_n$ DoF is allocated to boost the signal power for file $n$ and $M-K_n$ DoF is used to cancel the interferences from the $M-K_n$ nearest interfering helpers, where $K_n$ 
satisfies
\small{\begin{align}
K_n\in\mathcal M\triangleq\{1,2,\cdots,M\},\quad n\in\mathcal N.\label{eqn:DoF_allocation_PZF}
\end{align}}\normalsize
Thus, the adopted PZF receiver depends on parameter $\mathbf K\triangleq(K_n)_{n\in\mathcal N}$.
Let $c_i$ denote the index of the $i$-th nearest interfering helper. The PZF receive filter $\mathbf w_{0,n}^{\rm pzf}$ is the projection of the channel vector $\mathbf h_{\ell_{0,n},0}$ onto the subspace orthogonal to the one spanned by the channel vectors $\mathbf h_{c_1,0},\ \mathbf h_{c_2,0},\cdots,\mathbf h_{c_{M-K_n},0}$ of the $M-K_n$ canceled interferers $c_1,c_2,\cdots,c_{M-K_n}$. If the columns of an $M\times K_n$ matrix $\mathbf U_0$ form an orthonormal bases of the subspace, then the PZF receive filter is given by $\mathbf w_{0,n}^{\rm pzf}=\frac{\mathbf U_0\mathbf U_0^H\mathbf h_{\ell_{0,n},0}}{\|\mathbf U_0\mathbf U_0^H\mathbf h_{\ell_{0,n},0}\|_2}$. By applying this filter, the interferences from the $M-K_n$ nearest interfering helpers $c_1,c_2,\cdots,c_{M-K_n}$ are suppressed and the  corresponding SIR of $u_0$ is given by
\small{\begin{align}
\text{SIR}_{n}^{\text{pzf}}=\frac{H_{\ell_{0,n},0}^{\text{pzf}}d_{\ell_{0,n}}^{-\alpha}}{\sum_{i\in \Phi_h \setminus\{\ell_{0,n},c_1,c_2,\cdots,c_{M-K_n}\}}H_{i,0}^{\text{pzf}}d_i^{-\alpha}},\label{eqn:SIR_PZF}
\end{align}}\normalsize
where $H_{\ell_{0,n},0}^{\text{pzf}}\triangleq \|(\mathbf w_{0,n}^{\rm pzf})^H\mathbf h_{\ell_{0,n},0}\|_2^2$ denotes the fading power of the direct link and $H_{i,0}^{\text{pzf}}\triangleq \|(\mathbf w_{0,n}^{\rm pzf})^H\mathbf h_{i,0}\|_2^2$ denotes the fading power of the link between the caching helper $i$ and $u_0$. Here, $H_{\ell_{0,n},0}^{\text{pzf}}\dis \text{Gamma}(K_n,1)$ and $H_{i,0}^{\text{pzf}}\dis \text{Exp}(1)$ \cite{JindalTCOM11}. The PZF receiver is more general than the MRC receiver and reduces to the MRC receiver when $K_n=M$.
Note that the PZF receiver can efficiently cancel interference from the user's point of view, as a user is aware of its nearby interferers. Note that MISO transmission with a similar PZF beamformer may not achieve a SIR of the same distribution as the one in \eqref{eqn:SIR_PZF}, even with CSI at transmitters and information on nearby interferers of users (obtained via feadback from users). 

\subsection{Performance Metric}

When using the MRC receiver in the case of direct CSI, the transmission of file $n$ to $u_0$ is successful if $\text{SIR}_{n}^{\text{mrc}}\geq \tau$, where $\tau$ is the SIR threshold.
When using the PZF receiver in the case of local CSI, the transmission of file $n$ to $u_0$
is successful if $\text{SIR}_{n}^{\text{pzf}}\geq \tau$.
Therefore, the STPs of file $n\in\mathcal N$ requested by $u_0$ with the MRC receiver and the PZF receiver, denoted as $q_{M,n}^{\text{mrc}}(T_n)$ and $q_{M,n}^{\text{pzf}}(K_n,T_n)$ respectively, are given by
\small{\begin{align}
&q_{M,n}^{\text{mrc}}(T_n)\triangleq \Pr\left[\text{SIR}_{n}^{\text{mrc}}\geq\tau\right],\notag\\
&q_{M,n}^{\text{pzf}}(K_n,T_n)\triangleq \Pr\left[\text{SIR}_{n}^{\text{pzf}}>\tau\right],\notag
\end{align}}\normalsize
where $\text{SIR}_{n}^{\text{mrc}}$ and $\text{SIR}_{n}^{\text{pzf}}$ are given by \eqref{eqn:SIR_MRC} and \eqref{eqn:SIR_PZF}, respectively.
Note that the distributions of $\text{SIR}_{n}^{\text{mrc}}$ and $\text{SIR}_{n}^{\text{pzf}}$ depend on $T_n$  and $(T_n,K_n)$, respectively. 
Thus, we write $q_{M,n}^{\text{mrc}}$ and $q_{M,n}^{\text{pzf}}$ as functions of $T_n$ and $(T_n,K_n)$, respectively.
Requesters are mostly concerned about whether their desired files can be successfully received. Therefore, as in~\cite{Cui16TWC}, we consider the STP of a file randomly requested by $u_0$ as the network performance metric.
\footnote{Note that \cite{kuang2017random} considers random caching and MRT in MISO transmission for a large-scale cache-enabled network and studies a different performance metric. Therefore, the analysis and optimization framework in this paper is different from that in~\cite{kuang2017random}.}
According to the total probability theorem, the STPs of a file randomly requested by $u_0$ with the MRC receiver and the PZF receiver, denoted as $q_{M}^{\text{mrc}}(\mathbf T)$ and $q_M^{\text{pzf}}(\mathbf K,\mathbf T)$ respectively, are given by
\small{\begin{align}
&q_{M}^{\text{mrc}}(\mathbf T)\triangleq \Pr\left[\text{SIR}^{\rm mrc}\geq\tau\right]=\sum_{n\in\mathcal N}a_nq_{M,n}^{\text{mrc}}(T_n),\label{eqn:def_STP_MRC}\\
&q_M^{\text{pzf}}(\mathbf K,\mathbf T)\triangleq \Pr\left[\text{SIR}^{\rm pzf}\geq\tau\right]=\sum_{n\in\mathcal N}a_nq_{M,n}^{\text{pzf}}(K_n,T_n).\label{eqn:def_STP_PZF}
\end{align}}\normalsize

\section{Performance Analysis and Optimization for MRC Receiver}

In this section, we consider the performance analysis and optimization of the random caching design with the MRC receiver in the case of direct CSI. First, we analyze the STP in the general SIR threshold regime and the low SIR threshold regime, respectively. Then, we optimize the STP in these regimes.

\subsection{Performance analysis for MRC Receiver}


\subsubsection{Performance Analysis in General SIR Threshold Regime}
Note that different from \cite{Lin15TWC}, in the cache-enabled wireless network considered, 
there are two types of interferers, namely, i) interfering helpers storing the file requested by $u_0$ (which are farther than the serving helper of $u_0$), and ii) interfering helpers without the desired file of $u_0$ (which can be closer to $u_0$ than the serving helper of $u_0$). Different from \cite{Cui16TWC}, $u_0$ has $M$ receive antennas and performs  MRC to detect its desired signal. By carefully handling these two types of interferers and characterizing the impact of MRC, we obtain $q_{M,n}^{\text{mrc}}(T_n)$ using stochastic geometry. Substituting $q_{M,n}^{\text{mrc}}(T_n)$ into \eqref{eqn:def_STP_MRC}, we have the following result.
\begin{Thm}[STP with MRC Receiver in General SIR Threshold Regime]\label{Thm:STP_MRC_exact} 
The STP with the MRC receiver
 is given by $q_{M}^{\text{mrc}}(\mathbf T)=\sum_{n\in\mathcal N}a_nq_{M,n}^{\text{mrc}}(T_n)$, where
\small{\begin{align}
&q_{M,n}^{\text{mrc}}(T_n)=2\pi\lambda_h T_n\int_0^{\infty}x\exp\left(-\pi \lambda_h T_n x^2\right)\sum_{m=0}^{M-1}\frac{1}{m!}\sum_{k=0}^m\binom{m}{k}\tilde{\mathcal L}_{I}^k(T_n,x,x)\tilde{\mathcal L}_{I}^{m-k}(1-T_n,x,0){\rm d}x,\label{eqn:MRC_STP_file_n_exact}\\
&\mathcal L_{I}(T,x,y)=\exp\left(-\frac{2\pi\lambda_h T \tau^{\frac{2}{\alpha}}x^2}{\alpha}B'\left(\frac{2}{\alpha},1-\frac{2}{\alpha},\frac{1}{1+\tau(\frac{x}{y})^{\alpha}}\right)\right),\label{eqn:L_I_T_x_y}\\
&\tilde{\mathcal L}_{I}^k(T,x,y)=\mathcal L_{I}(T,x,y)\sum_{(b_j)_{j=1}^k\in\mathcal M_k}\frac{k!}{\prod_{j=1}^kb_j!}\prod_{j=1}^k\left(\frac{2\pi\lambda T\tau^{\frac{2}{\alpha}}x^2}{\alpha}B'\left(\frac{2}{\alpha}+1,j-\frac{2}{\alpha},\frac{1}{1+\tau(\frac{x}{y})^{\alpha}}\right)\right)^{b_j}.\label{eqn:L_I_k_T_x_y}
\end{align}}\normalsize
Here, 
$\mathcal{M}_{k}\triangleq\{(m_{j})_{j=1}^{k}|m_{j}\in\mathbb{N}^{0},\sum_{j=1}^{k}j\cdot m_{j}=k\}$, $\mathbb{N}^{0}$ denotes the set of nonnegative integers
and $B^{'}(a,b,z)\triangleq\int_{z}^{1}u^{a-1}(1-u)^{b-1}{\rm d}u$ ($0<z<1$) denotes the complementary incomplete beta function. 
\end{Thm}
\begin{IEEEproof}
Please refer to Appendix A.
\end{IEEEproof}

\begin{figure}[t]
\begin{center}
 \includegraphics[width=5.5cm]{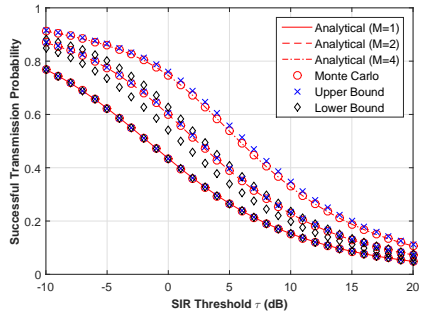}
  \end{center}
  \vspace{-6mm}
  \caption{\small{STP with the MRC receiver versus SIR threshold $\tau$.
  $N=5$, $C=3$, $\alpha=4$, $\lambda=10^{-3}$, $\mathbf T=[1,0.8,0.6,0.4,0.2]$, and $a_n=\frac{n^{-\gamma}}{\sum_{n\in \mathcal N}n^{-\gamma}}$ with $\gamma=1$.}}
  \vspace{-8mm}
\label{fig:exact_bounds}
\end{figure}

From Theorem~\ref{Thm:STP_MRC_exact}, we can see that the STP $q_{M}^{\text{mrc}}(\mathbf T)$ is an increasing function of the number of receive antennas $M$ at each user.
In particular, when increasing the number of receive antennas from $M-1$ to $M$, the increase of the STP of file $n$ is
\small{\begin{align}
&q_{M,n}^{\text{mrc}}(T_n)-q_{M-1,n}^{\text{mrc}}(T_n)\notag\\
&=\frac{2\pi\lambda_h  T_n}{(M-1)!}\int_0^{\infty}x\exp\left(-\pi \lambda_h T_n x^2\right)\sum_{k=0}^{M-1}\binom{M-1}{k}\tilde{\mathcal L}_{I}^{k}(T_n,x,x)\tilde{\mathcal L}_I^{M-1-k}(1-T_n,x,0){\rm d}x>0.\label{eqn:MRC_increse_STP_due_M}
\end{align}}\normalsize
From Theorem~\ref{Thm:STP_MRC_exact}, we also see that the impact of the physical layer parameters $\alpha$, $\tau$, $\lambda_h$ and the impact of the caching distribution $\mathbf T$ on $q_{M}^{\text{mrc}}(\mathbf T)$ are coupled in a very complex manner. Fig.~\ref{fig:exact_bounds} plots $q_M^{\text{mrc}}(\mathbf T)$ versus $\tau$ at different $M$. From Fig.~\ref{fig:exact_bounds}, we can see that each ``analytical'' curve (plotted using Theorem~\ref{Thm:STP_MRC_exact}) closely matches the corresponding ``Monte Carlo'' curve, verifying Theorem~\ref{Thm:STP_MRC_exact}. In addition, from Fig.~\ref{fig:exact_bounds}, we can see that $q_M^{\text{mrc}}(\mathbf T)$ increases with $M$. 

When $M=1$, by Theorem~\ref{Thm:STP_MRC_exact}, we can obtain a  simplified closed-form expression for $q^{\rm mrc}_{1}(\mathbf T)$.
\begin{Cor}[STP with MRC Receiver for $M=1$]\label{Cor:MRC_STP_M1}
For $M=1$, the STP with the MRC receiver is $q^{\rm mrc}_{1}(\mathbf T)=\sum_{n\in\mathcal N}a_nq^{\rm mrc}_{1,n}(T_n)$, where
\small{\begin{align}
q^{\rm mrc}_{1,n}(T_n)=&\frac{T_n}{c_{1,1}(1)T_n+c_{2,1}(1)}.\notag
\end{align}}\normalsize
Here, $c_{1,k}(x)$ and $c_{2,k}(x)$ are given by
\small{\begin{align}
c_{1,k}(x)\triangleq& \frac{2}{\alpha}\left(kx\tau\right)^{\frac{2}{\alpha}}\Bigg(B'\left(
\frac{2}{\alpha},1-\frac{2}{\alpha},\frac{1}{1+kx\tau}\right)-B\left(\frac{2}{\alpha},1-\frac{2}{\alpha}\right)\Bigg)+1,\label{eqn:c1}\\
c_{2,k}(x)\triangleq& \frac{2}{\alpha}\left(kx\tau\right)^{\frac{2}{\alpha}}
B\left(\frac{2}{\alpha},1-\frac{2}{\alpha}\right),\label{eqn:c2}
\end{align}}\normalsize
where $B(a,b)\triangleq\int_{0}^{1}u^{a-1}(1-u)^{b-1}{\rm d}u$ denotes the  beta function.
\end{Cor}

Note that Corollary~\ref{Cor:MRC_STP_M1} coincides with Corollary~4 of our previous work \cite{Cui16TWC}, which considers single-antenna users. From Corollary~\ref{Cor:MRC_STP_M1}, we can see that the impact of the physical layer parameters $\alpha$ and $\tau$ (captured by $c_{1,1}(1)$ and $c_{2,1}(1)$) and the impact of the caching distribution $\mathbf T$ on the STP $q^{\rm mrc}_1(\mathbf T)$ can be easily separated.
In addition, $q^{\rm mrc}_{1,n}(T_n)$ is a concave increasing function of $T_n$. This is because the average distance between a user requesting file $n$ and its serving helper decreases with $T_n$.



To facilitate the characterization of the STP $q_{M}^{\text{mrc}}(\mathbf T)$ for $M\geq 2$,
we next derive its closed-form upper and lower bounds, 
based on the upper and lower bounds on the incomplete gamma function, i.e.,  $\left(1-e^{-S_ab}\right)^a<\frac{\gamma(a,b)}{\Gamma(a)}<\left(1-e^{-b}\right)^a$ for $a>1$, where $S_a\triangleq\Gamma(a+1)^{-\frac{1}{a}}$ \cite{alzer1997some}. 
\begin{Lem}[Upper and Lower Bounds]\label{Lem:bounds_MRC}
The upper bound is given by $q_M^{\text{mrc},u}(\mathbf T)=\sum_{n\in\mathcal N}a_nq_{M,n}^{\text{mrc},u}(T_n)$, where
\small{\begin{align}
q_{M,n}^{\text{mrc},u}(T_n)&\triangleq 1-\sum_{k=0}^M\binom{M}{k}\frac{(-1)^kT_n}{c_{1,k}(S_M)T_n+c_{2,k}(S_M)},\notag
\end{align}}\normalsize
and the lower bound is given by $q_M^{\text{mrc},l}(\mathbf T)=\sum_{n\in\mathcal N}a_nq_{M,n}^{\text{mrc},l}(T_n)$, where
\small{\begin{align}
q_{M,n}^{\text{mrc},l}(T_n)&\triangleq 1-\sum_{k=0}^M\binom{M}{k}\frac{(-1)^kT_n}{c_{1,k}(1)T_n+c_{2,k}(1)}.\notag
\end{align}}\normalsize
Here, $c_{1,k}(x)$ and $c_{2,k}(x)$ are given by \eqref{eqn:c1} and \eqref{eqn:c2}, respectively.
\end{Lem}
\begin{IEEEproof}
Please refer to Appendix B.
\end{IEEEproof}

Note that the upper bound $q^{\text{mrc},u}_M(\mathbf T)$ and the lower bound $q^{\text{mrc},l}_M(\mathbf T)$ coincide when $M=1$, i.e., $q_M^{\text{mrc},u}(\mathbf T)=q_M^{\text{mrc},l}(\mathbf T)=q_M^{\text{mrc}}(\mathbf T)$, implying that the two bounds are tight at $M=1$. 
Similarly, for $M\geq 2$, the impact of the physical layer parameters $\alpha$ and $\tau$ (captured by $c_{1,k}(k)$, $c_{2,k}(k)$, $k\in\{0,\cdots,M\}$) and the impact of the caching distribution $\mathbf T$ on $q^{\text{mrc},u}_M(\mathbf T)$ and $q^{\text{mrc},l}_M(\mathbf T)$ can be easily separated.

Fig.~\ref{fig:exact_bounds} plots $q_M^{\text{mrc},u}(\mathbf T)$, $q_M^{\text{mrc},l}(\mathbf T)$ and $q_M^{\text{mrc}}(\mathbf T)$ versus $\tau$ at different $M$. From Fig.~\ref{fig:exact_bounds}, we can see that when $M=1$, $q_M^{\text{mrc}}(\mathbf T)$, $q_M^{\text{mrc},u}(\mathbf T)$ and $q_M^{\text{mrc},l}(\mathbf T)$ coincide; when $M\geq2$, $q_M^{\text{mrc},u}(\mathbf T)$ and $q_M^{\text{mrc},l}(\mathbf T)$ bound $q_M^{\text{mrc}}(\mathbf T)$ from above and below, respectively. In addition, $q_M^{\text{mrc},u}(\mathbf T)$ tightly matches  $q_M^{\text{mrc}}(\mathbf T)$ over the entire range of SIR threshold $\tau$ of interest,
demonstrating that $q_M^{\text{mrc},u}(\mathbf T)$ can serve as a good approximation for $q_M^{\text{mrc}}(\mathbf T)$.

\subsubsection{Performance Analysis in Low SIR Threshold Regime}

Although the impacts of the physical layer parameters and the caching distribution can be separated in $q_M^{\text{mrc},u}(\mathbf T)$, how these parameters affect $q_M^{\text{mrc},u}(\mathbf T)$ is still not clear.
To further obtain insights,
we analyze the outage probability $\overline{q}^{\text{mrc}}_{M}(\mathbf T) \triangleq \Pr\left[\text{SIR}^{\rm mrc}<\tau\right]= 1-q_M^{\text{mrc}}(\mathbf T)$ in the low SIR threshold regime, i.e., $\tau\to 0$.\footnote{The downlink transmission in the LTE system supports an SINR about $-7$ dB. 
Thus, $\tau$ can be very small and the asymptotic analysis is applicable in certain scenarios~\cite{zhang14}.} 
\begin{Lem}[Outage Probability with MRC Receiver in Low SIR Threshold Regime]\label{Lem:STP_low_SIR}
As $\tau\to 0$, $\overline{q}_{M}^{\text{mrc}}(\mathbf T)=\overline{q}^{\text{mrc}}_{M,0}(\mathbf T)+o(\tau^{\frac{2}{\alpha}})$, where
\small{\begin{align}
&\overline{q}^{\text{mrc}}_{M,0}(\mathbf T)=\sum_{n\in\mathcal N}a_n\overline{q}^{\text{mrc}}_{M,0,n}(T_n),\label{eqn:MRC_OP}\\
&\overline{q}^{\text{mrc}}_{M,0,n}(T_n)=\tau^{\frac{2}{\alpha}}\frac{2}{\alpha}\left(\frac{1}{T_n}-1\right)\sum_{m=M}^{\infty}B\left(\frac{2}{\alpha}+1,m-\frac{2}{\alpha}\right).\label{eqn:MRC_OP_n}
\end{align}}\normalsize
\end{Lem}
\begin{IEEEproof}
Please refer to Appendix C.
\end{IEEEproof}

\begin{figure}[t]
\begin{center}
\includegraphics[width=5.5cm]{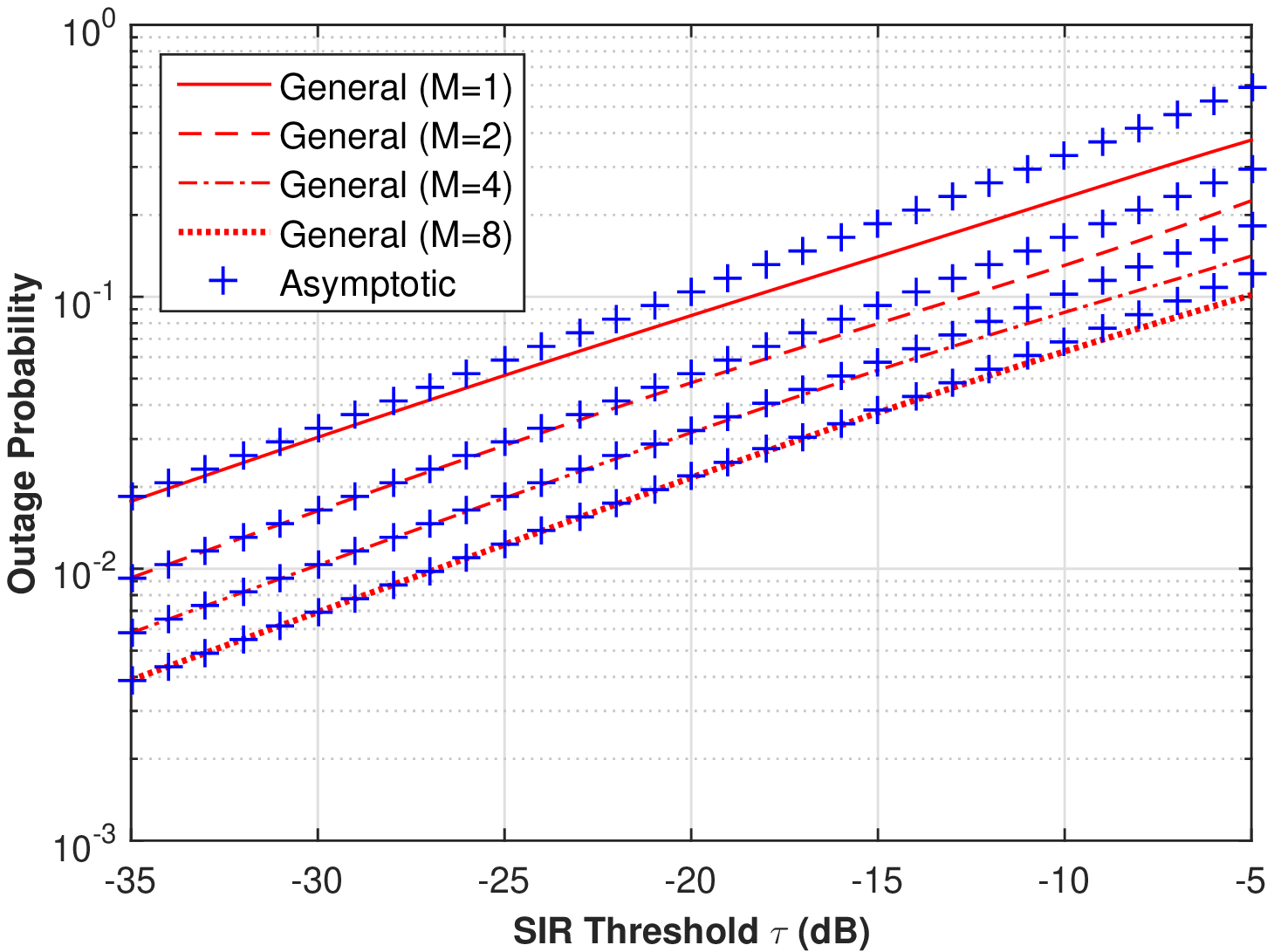}
\end{center}
\vspace{-6mm}
\caption{\small{Outage probability with the MRC receiver versus SIR threshold $\tau$.
$N=5$, $C=3$, $\alpha=4$, $\lambda=10^{-3}$, $\mathbf T=[1,0.8,0.6,0.4,0.2]$, and $a_n=\frac{n^{-\gamma}}{\sum_{n\in \mathcal N}n^{-\gamma}}$ with $\gamma=1$.}}
\vspace{-8mm}
\label{fig:MRC_STP_asymp}
\end{figure}

First, we introduce the order gain of the outage probability, defined as the exponent of the outage probability as SIR threshold $\tau$ decreases to $0$~\cite{zhang14}, i.e., $d^{\rm mrc} \triangleq \lim_{\tau\to 0}\frac{\text{log}\Pr[\text{SIR}^{\text{mrc}}<\tau]}{\text{log}\tau}$.
Then, we define the coefficient of the asymptotic outage probability, i.e., $e^{\rm mrc}\triangleq \lim_{\tau\to 0}\frac{\Pr[\text{SIR}^{\text{mrc}}<\tau]}{\tau^{d^{\rm mrc}}}$. Leveraging its order gain and the coefficient, we now characterize the key behavior of the outage probability $\overline{q}_M^{\text{mrc}}(\mathbf T)$ in the low SIR threshold regime based on Lemma~\ref{Lem:STP_low_SIR}.
Specifically, the order gain $d^{\rm mrc}$ is $\frac{2}{\alpha}$, which does not depend on the number of receive antennas $M$ and the caching distribution $\mathbf T$.
The coefficient $e^{\rm mrc}$ is affected by $M$ and $\mathbf T$.
Specifically, when increasing the number of receive antennas from $M-1$ to $M$, the decrease of the coefficient $e^{\rm mrc}$  is
\small{\begin{align}
&\frac{2}{\alpha}\sum_{n\in\mathcal N}a_n\left(\frac{1}{T_n}-1\right)\sum_{m=M-1}^{\infty}B\left(\frac{2}{\alpha}+1,m-\frac{2}{\alpha}\right)-\frac{2}{\alpha}\sum_{n\in\mathcal N}a_n\left(\frac{1}{T_n}-1\right)\sum_{m=M}^{\infty}B\left(\frac{2}{\alpha}+1,m-\frac{2}{\alpha}\right)\notag\\
&=\frac{2}{\alpha}B\left(\frac{2}{\alpha}+1,M-1-\frac{2}{\alpha}\right)\sum_{n\in\mathcal N}a_n\left(\frac{1}{T_n}-1\right)>0.\notag 
\end{align}}\normalsize
Fig.~\ref{fig:MRC_STP_asymp} plots $\overline{q}^{\text{mrc}}_{M}(\mathbf T)$ and $\overline{q}^{\text{mrc}}_{M,0}(\mathbf T)$ versus the SIR threshold $\tau$ in the low SIR threshold regime. We can see from Fig.~\ref{fig:MRC_STP_asymp} that when $\tau$ decreases, the gap between each ``General'' curve $\overline{q}^{\text{mrc}}_{M}(\mathbf T)$, which is plotted using Theorem~\ref{Thm:STP_MRC_exact}, and the corresponding ``Asymptotic'' curve $\overline{q}^{\text{mrc}}_{M,0}(\mathbf T)$, which is plotted using Lemma~\ref{Lem:STP_low_SIR}, decreases, verifying Lemma~\ref{Lem:STP_low_SIR}.
In addition, from Fig.~\ref{fig:MRC_STP_asymp}, we can see that ``Asymptotic'' curves with different $M$ have the same slope (implying the same order gain), and there is a shift between two  ``Asymptotic'' curves with different $M$ (implying different coefficients).

\subsection{Performance Optimization for MRC Receiver}


\subsubsection{Performance Optimization in General SIR Threshold Regime}
In the general SIR threshold regime, we would like to optimize $\mathbf T$ to maximize the STP with the MRC receiver.
Recall that the closed-form upper bound $q_M^{\text{mrc},u}(\mathbf T)$
provides a good approximation for $q_M^{\text{mrc}}(\mathbf T)$, as shown in Fig.~\ref{fig:exact_bounds}. In addition, $q_M^{\text{mrc},u}(\mathbf T)$ has a much simpler form than $q_M^{\text{mrc}}(\mathbf T)$. 
Thus, in the following, we maximize $q_M^{\text{mrc},u}(\mathbf T)$ 
instead of directly maximizing $q_M^{\text{mrc}}(\mathbf T)$.

\begin{Prob}[Performance Optimization for MRC Receiver]\label{Prob:MRT_ub}
\small{\begin{align}
\max_{\mathbf T}& \quad q_M^{\text{mrc},u}(\mathbf T) \notag\\
s.t.& \quad \eqref{eqn:cache_interval_constr}, \eqref{eqn:cache_size_constr}. \notag
\end{align}}\normalsize
\end{Prob}

When $M=1$, Problem~\ref{Prob:MRT_ub} is a convex optimization problem, and its closed-form optimal solution is given in Theorem~$4$ of our previous work \cite{Cui16TWC}. When $M\geq 2$,  Problem~\ref{Prob:MRT_ub} is in general a non-convex optimization problem with a differentiable non-convex objective function and a convex constraint set.
The gradient projection method can be applied to obtain a stationary point of Problem~\ref{Prob:MRT_ub}.\footnote{Note that a stationary point is a point that satisfies the necessary optimality conditions of a non-convex optimization problem, and it is the classic goal in the design of iterative algorithms for non-convex optimization problems.}
However, the rate of convergence of the gradient projection algorithm is strongly dependent on the choices of the stepsize and initial point. If they are chosen improperly, it may take a large number of iterations to meet some convergence criterion~\cite{Zitian17}.
To address this issue, in the following, we propose a more efficient algorithm to obtain a stationary point of Problem~\ref{Prob:MRT_ub}.

First, we can see that Problem~\ref{Prob:MRT_ub} is equivalent to the minimization of the difference of two convex functions $R_o(\mathbf T)\triangleq-\sum_{n\in\mathcal N}a_n\sum_{i=0}^{\lceil\frac{M}{2}\rceil-1}\binom{M}{2i+1}\frac{T_n}{C_{1,2i+1}(S_M)T_n+C_{2,2i+1}(S_M)}$ and $R_e(\mathbf T)\triangleq-\sum_{n\in\mathcal N}a_n\sum_{i=0}^{\lfloor\frac{M}{2}\rfloor}\binom{M}{2i}\frac{T_n}{C_{1,2i}(S_M)T_n+C_{2,2i}(S_M)}$ subject to the constraints in \eqref{eqn:cache_interval_constr} and \eqref{eqn:cache_size_constr}, which is given below.
\begin{Prob}[Equivalent Problem of Problem \ref{Prob:MRT_ub}]\label{Prob:MRT_ub_equivalent}
\small{\begin{align}
\min_{\mathbf T}& \quad R_o(\mathbf T)-R_e(\mathbf T) \notag\\
s.t.& \quad \eqref{eqn:cache_interval_constr}, \eqref{eqn:cache_size_constr}.\notag
\end{align}}\normalsize
\end{Prob}

Problem~\ref{Prob:MRT_ub_equivalent} is a DC programming problem~\cite{Horst1999}.\footnote{An optimization problem is called a DC programming problem if its variables are restricted to a convex set and its objective function and its inequality constraint functions are DC functions~\cite{Horst1999}.}
Note that constructing an algorithm to find a globally optimal solution of a DC programming problem is in general an open problem. We use CCCP to obtain a stationary point of Problem~\ref{Prob:MRT_ub_equivalent} \cite{Lanckriet09}.
The main idea of CCCP is to approximate the objective DC function 
by replacing the second term (i.e., $R_e(\mathbf T)$) with its first order Taylor expansion, and then solve a sequence of convex problems successively. Specifically, at iteration $t$ of CCCP,  we have the following problem.
\begin{Prob}[Problem at Iteration $t$ of CCCP]\label{Prob:MRT_ub_equivalent_sub}
\small{\begin{align}
\min_{\mathbf T}& \quad R_o(\mathbf T)+\sum_{n\in\mathcal N}a_nf_e(T_n^{\dagger}(t-1))T_n \notag\\
s.t.& \quad \eqref{eqn:cache_interval_constr}, \eqref{eqn:cache_size_constr},\notag
\end{align}}\normalsize
where $f_e(x)\triangleq -\frac{1}{a_n}\frac{\partial R_e(\mathbf T)}{\partial T_n}\Big{|}_{T_n=x}=\sum_{i=0}^{\lfloor\frac{M}{2}\rfloor}\binom{M}{2i}\frac{C_{2,2i}(S_M)}{(C_{1,2i}(S_M)x+C_{2,2i}(S_M))^2}$,
and $\mathbf T^{\dagger}(t)$ denotes the optimal solution of the problem at iteration $t$ of CCCP.
\end{Prob}

It can be easily seen that Problem~\ref{Prob:MRT_ub_equivalent_sub} is a convex optimization problem and Slater's condition is satisfied, implying that strong duality holds. Using KKT conditions, we can solve Problem~\ref{Prob:MRT_ub_equivalent_sub}.
\begin{Lem}[Optimal Solution of Problem~\ref{Prob:MRT_ub_equivalent_sub}]\label{Lem:opt_subprob}
The optimal solution $\mathbf T^{\dagger}(t)$ of Problem~\ref{Prob:MRT_ub_equivalent_sub} is 
\small{\begin{align}
&T_n^{\dagger}(t)=\begin{cases}
0, \hspace{40mm} f_o(0)<f_e(T_n^{\dagger}(t-1))+\frac{v^{\dagger}(t)}{a_n}\\
1, \hspace{40mm} f_o(1)>f_e(T_n^{\dagger}(t-1))+\frac{v^{\dagger}(t)}{a_n}\\
x(T_n^{\dagger}(t-1),a_n,v^{\dagger}(t)), \quad\quad\ {\rm otherwise}
\end{cases},\quad n\in\mathcal N,\notag
\end{align}}\normalsize
where $f_o(x)\triangleq -\frac{1}{a_n}\frac{\partial R_o(\mathbf T)}{\partial T_n}\Big{|}_{T_n=x}= \sum_{i=0}^{\lceil\frac{M}{2}\rceil-1}\binom{M}{2i+1}\frac{C_{2,2i+1}(S_M)}{(C_{1,2i+1}(S_M)x+C_{2,2i+1}(S_M))^2}$,  $x(T_n^{\dagger}(t-1),a_n,v^{\dagger}(t))$ denotes the solution of equation $f_o(x)=f_e(T_n^{\dagger}(t-1))+\frac{v^{\dagger}(t)}{a_n}$, and $v^{\dagger}(t)$ satisfies $\sum\limits_{n\in\mathcal N}T_n^{\dagger}(t)=C$.
\end{Lem}
\begin{IEEEproof}
Please refer to Appendix D.
\end{IEEEproof}


\vspace{-4mm}
\begin{algorithm} \caption{Stationary Point of Problem~\ref{Prob:MRT_ub_equivalent} based on CCCP}
\small{\begin{algorithmic}[1]
\STATE Initialize $t=1$, $T_n^{\dagger}(1)=\frac{C}{N}$ for $n\in\mathcal N$, $\epsilon=10^{-4}$.
\STATE \textbf{repeat}
\STATE Obtain $\mathbf T^{\dagger}(t+1)$ according to Lemma~\ref{Lem:opt_subprob}.
\STATE Set $t=t+1$. 
\STATE \textbf{until} $q_M^{\text{mrc},u}(\mathbf T^{\dagger}(t))-q_M^{\text{mrc},u}(\mathbf T^{\dagger}(t-1))<\epsilon$
\end{algorithmic}}\normalsize\label{alg:MRC_CCCP}
\end{algorithm}
\vspace{-6mm}

Since $f_o(x)$ is a decreasing function, for given $v^{\dagger}(t)$ and $ T_n^{\dagger}(t-1)$, $x(T_n^{\dagger}(t-1),a_n,v^{\dagger}(t))$ can be efficiently obtained by bisection search. In addition, since $\sum_{n\in\mathcal N}T_n^{\dagger}(t)$ decreases with $v^{\dagger}(t)$, $v^{\dagger}(t)$ can also be efficiently obtained by bisection search. The details for solving Problem~\ref{Prob:MRT_ub_equivalent} using CCCP are summarized in Algorithm~\ref{alg:MRC_CCCP}. Different from the gradient projection method, Algorithm~\ref{alg:MRC_CCCP} does not rely on any stepsize. From~\cite{Lanckriet09}, we know that Algorithm~\ref{alg:MRC_CCCP} converges to a stationary point of Problem~\ref{Prob:MRT_ub_equivalent}, denoted by $\mathbf T^{\dagger}$.  Thus, Algorithm~\ref{alg:MRC_CCCP} may have more robust convergence performance than the gradient projection method.
By analyzing structural properties of the stationary point $\mathbf T^{\dagger}$ of Problem~\ref{Prob:MRT_ub_equivalent} obtained by Algorithm~\ref{alg:MRC_CCCP}, we have the following result.

\begin{Lem}[Property of Stationary Point]\label{Lem:property_solution_MRC}
The stationary point $\mathbf T^{\dagger}$ of Problem~\ref{Prob:MRT_ub_equivalent} obtained by Algorithm~\ref{alg:MRC_CCCP} satisfies $T_1^{\dagger}\geq\cdots\geq T_n^{\dagger}$.
\end{Lem}
\begin{IEEEproof}
Please refer to Appendix E.
\end{IEEEproof}

\begin{figure}[t]
\begin{center}
\includegraphics[width=5.5cm]{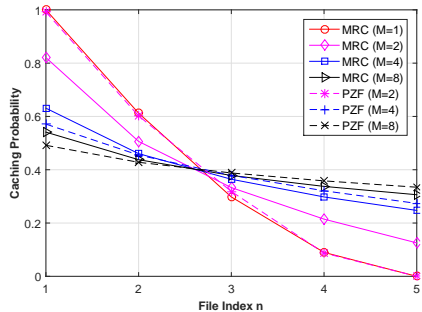}
\end{center}
\vspace{-6mm}
\caption{\small{Optimized caching probability $T_n^{\dagger}$ versus file index $n$.
$N=5$, $C=2$, $\alpha=4$, $\tau=0.5$, $\lambda=10^{-3}$, and $a_n=\frac{n^{-\gamma}}{\sum_{n\in \mathcal N}n^{-\gamma}}$ with $\gamma=0.4$.}}
\vspace{-8mm}
\label{fig:opt_MRT_ub_Tn_vs_n}
\end{figure}

Lemma~\ref{Lem:property_solution_MRC} reveals that files of higher popularity get more storage resources.
Fig.~\ref{fig:opt_MRT_ub_Tn_vs_n} shows $\mathbf T^{\dagger}$ at different $M$. From Fig.~\ref{fig:opt_MRT_ub_Tn_vs_n}, we can see that files of higher popularity get more storage resources, verifying Lemma~\ref{Lem:property_solution_MRC}. In addition, $T_n^{\dagger}$ decreases with $n$ more slowly when $M$ is larger, i.e., $\mathbf T^{\dagger}$ becomes more flat when $M$ is larger.
This is because when $M$ increases, $u_0$ can 
receive its desired file from the serving helper at a larger distance from $u_0$. Storing more different files (corresponding to a more flat caching distribution) can increase the STP with the MRC receiver.

\subsubsection{Performance Optimization in Low SIR Threshold Regime}

In this part, we consider the minimization of the asymptotic outage probability $\overline{q}^{\text{mrc}}_{M,0}(\mathbf T)$ (i.e., the maximization of the asymptotic STP) in the low SIR threshold regime.
\begin{Prob}[Performance Optimization for MRC Receiver in Low SIR Threshold Regime]\label{Prob:MRC_asymp}
\small{\begin{align}
\min_{\mathbf T}& \quad \overline{q}^{\text{mrc}}_{M,0}(\mathbf T) \notag\\
s.t.& \quad \eqref{eqn:cache_interval_constr}, \eqref{eqn:cache_size_constr}. \notag
\end{align}}\normalsize
Let $\mathbf T_0^*\triangleq (T_{0,n}^*)_{n\in\mathcal N}$ denote the optimal solution.
\end{Prob}

Problem~\ref{Prob:MRC_asymp} is a convex optimization problem and Slater's condition is satisfied, implying that strong duality holds. Using KKT conditions, we can solve Problem~\ref{Prob:MRC_asymp}.
\begin{Lem}[Optimal Solution of Problem~\ref{Prob:MRC_asymp}]\label{Lem:opt_asymp_MRC}
The optimal solution $\mathbf T_0^*$ of Problem~\ref{Prob:MRC_asymp} is 
\small{\begin{align}
T_{0,n}^*=\min\left\{\sqrt{\frac{a_n}{\nu_0}},1\right\},\notag
\end{align}}\normalsize
where $\nu_0$ satisfies $\sum_{n\in\mathcal N}\min\left\{\sqrt{\frac{a_n}{\nu_0}},1\right\} = C$.
\end{Lem}
\begin{IEEEproof}
Lemma~\ref{Lem:opt_asymp_MRC} can be proved in a similar way to Lemma~\ref{Lem:opt_subprob}. We omit the details due to page limitation.
\end{IEEEproof}

\begin{figure}[t]
\begin{center}
\includegraphics[width=5.5cm]{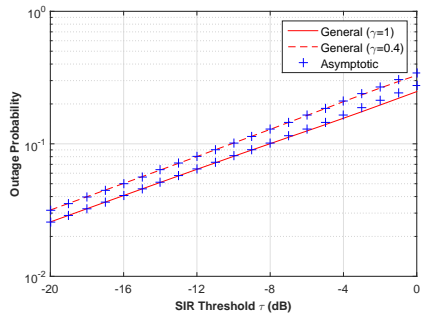}
\end{center}
\vspace{-6 mm}
\caption{\small{Outage probabilities with the MRC receiver $\overline{q}_M^{\rm mrc}(\mathbf T^{\dagger})$ and $\overline{q}_M^{\rm mrc}(\mathbf T_0^{*})$ versus SIR threshold $\tau$.
$N=5$, $C=3$, $M=4$, $\alpha=4$, $\lambda=10^{-3}$, and $a_n=\frac{n^{-\gamma}}{\sum_{n\in \mathcal N}n^{-\gamma}}$.}}
\vspace{-8 mm}
\label{fig:MRC_opt_general_vs_asymp}
\end{figure}

Note that $\nu_0$ can be efficiently obtained by bisection search.
From Lemma~\ref{Lem:opt_asymp_MRC}, we can see that
$\mathbf T_0^*$ is not affected by the number of receive antennas $M$.
Fig.~\ref{fig:MRC_opt_general_vs_asymp} plots $\overline{q}^{\text{mrc}}_{M}(\mathbf T^{\dagger})$ and $\overline{q}^{\text{mrc}}_{M}(\mathbf T_0^*)$ versus the SIR threshold $\tau$ at different Zipf exponent $\gamma$.
From Fig.~\ref{fig:MRC_opt_general_vs_asymp}, we can see  that when $\tau$ decreases,
the gap between each ``General'' curve $\overline{q}^{\text{mrc}}_{M}(\mathbf T^{\dagger})$ and the corresponding ``Asymptotic''  curve $\overline{q}^{\text{mrc}}_{M}(\mathbf T_0^*)$ decreases. Thus, Fig.~\ref{fig:MRC_opt_general_vs_asymp} verifies Lemma~\ref{Lem:opt_asymp_MRC} and the optimality of $\mathbf T^{\dagger}$ in the low SIR threshold regime.

\section{Performance Analysis and Optimization for PZF Receiver}

In this section, we consider the performance analysis and optimization of the random caching design with the PZF receiver in the case of local CSI. First, we analyze the STP in the general SIR threshold regime. Then, we optimize the STP.

\subsection{Performance Analysis for PZF receiver}



Note that when requesting file $n$, $u_0$ can cancel the interferences from the $M-K_n$ nearest interfering helpers, using the PZF receiver parameterized by $\mathbf K$. Specifically, when $\ell_{0,n}>M-K_n$, the interferences from the helpers in $\{1,2,\cdots, M-K_n\}$ are canceled. When $\ell_{0,n}\leq M-K_n$, the interferences from the helpers in $\{1,2,\cdots,M-K_n+1\}\setminus\{\ell_{0,n}\}$ are canceled. By carefully handling these two scenarios, we can obtain $q^{\text{pzf}}_{M,n}(K_n,T_n)$ using stochastic geometry. Substituting $q_{M,n}^{\text{pzf}}(K_n,T_n)$ into \eqref{eqn:SIR_PZF}, we have the following result.

\begin{Thm}[STP with PZF Receiver in General SIR Threshold Regime]\label{Thm:STP_PZF_exact}
The STP with the PZF receiver parameterized by $\mathbf K$ is given by $q^{\text{pzf}}_M(\mathbf K,\mathbf T)=\sum_{n\in\mathcal N}a_nq^{\text{pzf}}_{M,n}(K_n,T_n)$, where
\small{\begin{align}
&q_{M,n}^{\text{pzf}}(K_n,T_n)\notag\\
&=\begin{cases}
(1-T_n)^{M-K_n} \int_{0}^{\infty}\int_{0}^{x}g_{d_{\ell_{0,n}},d_{M-K_n}}(T_n,x,y) \sum_{m=0}^{K_n-1}\frac{1}{m!}\sum_{k=0}^m\binom{m}{k}\tilde{\mathcal L}_{I}^k(T_n,x,x)\tilde{\mathcal L}_{I}^{m-k}(1-T_n,x,y){\rm d}y{\rm d}x\\
\quad+\sum_{m=1}^{M-K_n}T_n(1-T_n)^{m-1}\int_{0}^{\infty}\int_{0}^{y}f_{d_m,d_{M-K_n+1}}(x,y)\sum_{k=0}^{K_n-1}\frac{\tilde{\mathcal L}_{I}^k(1,x,y)}{k!}{\rm d}x {\rm d}y, \hfill K_n<M\\ 
q_{M,n}^{\text{mrc}}(T_n),  \hfill K_n=M
\end{cases}\notag\\
&g_{d_{\ell_{0,n}},d_{j}}(T_n,x,y)=\frac{4\pi^{j+1}\lambda_h^{j+1} T_n x y^{2j-1}}{(j-1)!}\exp\left(-\pi\lambda_h T_nx^2-\pi\lambda_h(1-T_n)y^2\right),\label{eqn:g_j_T_x_y}\\
&f_{d_i,d_j}(x,y)=\frac{4\pi^j\lambda_h^jx^{2i-1}y(y^2-x^2)^{j-i-1}}{(i-1)!(j-i-1)!}\exp(-\pi\lambda_h y^2).\label{eqn:f_i_j_x_y}
\end{align}}\normalsize
Here, $q_{M,n}^{\text{mrc}}(T_n)$ is given by \eqref{eqn:MRC_STP_file_n_exact} and $\tilde{\mathcal L}_{I}^k(T,x,y)$ is given by \eqref{eqn:L_I_k_T_x_y}.
\end{Thm}
\begin{IEEEproof}
Please refer to Appendix F.
\end{IEEEproof}

\begin{figure}[t]
\begin{center}
 \includegraphics[width=5.5cm]{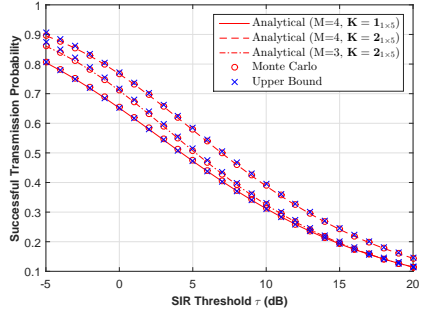}
  \end{center}
  \vspace{-6mm}
  \caption{\small{STP with the PZF receiver versus SIR threshold $\tau$.
  $N=5$, $C=3$, $\alpha=4$, $\lambda=10^{-3}$, $L=3$, $\mathbf T=[1,0.8,0.6,0.4,0.2]$,  and $a_n=\frac{n^{-\gamma}}{\sum_{n\in \mathcal N}n^{-\gamma}}$ with $\gamma=1$.}}
  \vspace{-8mm}
\label{fig:PZF_exact_bounds}
\end{figure}

Note that when $K_n=M$ for all $n\in\mathcal N$, Theorem~\ref{Thm:STP_PZF_exact} reduces to Theorem~\ref{Thm:STP_MRC_exact}, also verifying that the MRC receiver is a special case of the PZF receiver parameterized by $\mathbf K$.
From Theorem~\ref{Thm:STP_PZF_exact}, we can see that the STP $q_M^{\text{pzf}}(\mathbf K,\mathbf T)$ is an increasing function of the number of receive antennas $M$ at each user.
In particular, for $K_n=M$, when increasing the number of receive antennas from $M-1$ to $M$ and the number of DoF for file $n$ from $K_n-1$ to $K_n$, the increase of the STP of file $n$ is given by \eqref{eqn:MRC_increse_STP_due_M}; for $K_n<M$, when increasing the number of receive antennas from $M-1$ to $M$ and the number of DoF for file $n$ from $K_n-1$ to $K_n$, the increase of the STP of file $n$ is given by
\small{\begin{align}
&q^{\text{pzf}}_{M,n}(K_n,T_n)-q^{\text{pzf}}_{M-1,n}(K_n-1,T_n)=\sum_{m=1}^{M-K_n}\frac{T_n(1-T_n)^{m-1}}{(K_n-1)!}\int_{0}^{\infty}\int_{0}^{y}f_{d_m,d_{M-K_n+1}}(x,y)\tilde{\mathcal L}_{I}^{K_n-1}(1,x,y){\rm d}x {\rm d}y\notag\\
&+\frac{(1-T_n)^{M-K_n}}{(K_n-1)!}\int_{0}^{\infty}\int_{0}^{x}g_{d_{\ell_{0,n}},d_{M-K_n}}(T_n,x,y) \sum_{k=0}^{K_n-1}\binom{K_n-1}{k}\tilde{\mathcal L}_{I}^k(T_n,x,x)\tilde{\mathcal L}_{I}^{K_n-1-k}(1-T_n,x,y){\rm d}y{\rm d}x>0.\notag
\end{align}}\normalsize
In addition, from Theorem~\ref{Thm:STP_PZF_exact}, we see that
the impact of the physical layer parameters $\alpha$, $\tau$, $\lambda_h$, the impact of the caching distribution $\mathbf T$ and the impact of the DoF allocation $\mathbf K$ on $q_M^{\text{pzf}}(\mathbf K,\mathbf T)$ are coupled in a very complex manner; the impact of the caching distribution $\mathbf T$ on $q_M^{\text{pzf}}(\mathbf K,\mathbf T)$ is not clear. Fig.~\ref{fig:PZF_exact_bounds} plots $q^{\text{pzf}}_M(\mathbf K,\mathbf T)$ versus $\tau$ at different $M$ and $\mathbf K$. From Fig.~\ref{fig:PZF_exact_bounds}, we can see that each ``analytical'' curve closely matches the corresponding ``Monte Carlo'' curve. Thus, Fig.~\ref{fig:PZF_exact_bounds} verifies Theorem~\ref{Thm:STP_PZF_exact}. In addition, from Fig.~\ref{fig:PZF_exact_bounds}, we can see that $q^{\text{pzf}}_M(\mathbf K,\mathbf T)$ increases with $M$. 


To facilitate the characterization of the STP $q^{\text{pzf}}_M(\mathbf K,\mathbf T)$ and decouple the impacts of the caching distribution $\mathbf T$ and the physical layer parameters on $q^{\text{pzf}}_M(\mathbf K,\mathbf T)$,
we then derive a tractable  upper bound on $q^{\text{pzf}}_M(\mathbf K,\mathbf T)$, 
based on $\Pr\left[\text{SIR}_n^{\text{pzf}}>\tau|\ell_{0,n}=m\right]>\Pr\left[\text{SIR}_n^{\text{pzf}}>\tau|\ell_{0,n}=k\right]$ for any $k>m$ and the lower bound on the incomplete gamma function, i.e., $\left(1-e^{-S_ab}\right)^a<\frac{\gamma(a,b)}{\Gamma(a)}$ \cite{alzer1997some}. Note that the upper bound is parameterized by $L\in\{1,2,\cdots\}$. A larger value of $L$ corresponds to a tighter upper bound and higher computation complexity for calculating the upper bound.
The upper bound $q^{\text{pzf},u}_M(\mathbf K,\mathbf T)$ on $q^{\text{pzf}}_M(\mathbf K,\mathbf T)$ is given below.
\begin{Lem}[Upper Bound]\label{Lem:upper_bound_PZF}
The upper bound parameterized by $L\in\{1,2,\cdots\}$ is given by $q^{\text{pzf},u}_M(\mathbf K,\mathbf T)=\sum_{n\in\mathcal N}a_nq^{\text{pzf},u}_{M,n}(K_n,T_n)$, where
\small{\begin{align}
&q^{\text{pzf},u}_{M,n}(K_n,T_n)=
\sum_{m=1}^{M-K_n+L-1}T_n(1-T_n)^{m-1}R_{M,K_n,m}+(1-T_n)^{M-K_n+L-1}R_{M,K_n,M-K_n+L},\label{eqn:q_n_U_T_K}\\
&\hspace{-2mm}R_{M,K,m}
=\begin{cases}
\int_{0}^{\infty}\int_{0}^{y}f_{d_m,d_{M-K+1}}(x,y)\sum_{k=1}^{K}(-1)^{k+1}\mathcal L_{I}(1,(kS_K)^{\frac{1}{\alpha}}x,y){\rm d}x {\rm d}y, \quad m\in\{1,\cdots,M-K\}\\
\int_0^{\infty}h_{d_{M-K+1}}(x)\sum_{k=1}^{K}(-1)^{k+1}\mathcal L_{I}(1,(kS_K)^{\frac{1}{\alpha}}x,x){\rm d}x, \hspace{20mm} m=M-K+1\\
\int_{0}^{\infty}\int_{0}^{x}\int_{y}^{x}\cdots\int_{y}^{x}f_{d_{M-K+1},d_m}(y,x)
\left(\prod_{i=1}^{m-M+K-2}\frac{2r_i}{x^2-y^2}\right)\sum_{k=1}^{K}\binom{K}{k}(-1)^{k+1}\\
\quad\times\mathcal L_{I}(1,(kS_K)^{\frac{1}{\alpha}}x,x)\prod_{i=1}^{m-M+K-2}\frac{1}{1+kS_K\tau \left(\frac{x}{r_i}\right)^{\alpha}}\\
\quad\times\frac{1}{1+kS_K\tau \left(\frac{x}{y}\right)^{\alpha}}{\rm d}r_1\cdots{\rm d}r_{m-M+K-2}{\rm d}y{\rm d}x, \hspace{10mm}  m\in\{M-K+2,\cdots,M-K+L\}
\end{cases}.
\label{eqn:R_M_K_m}
\end{align}}\normalsize
Here, $h_{d_j}(x)\triangleq \frac{2\pi^j\lambda^jx^{2j-1}}{(j-1)!}\exp\left(-\pi\lambda x^2\right)$, 
$\tilde{\mathcal L}_{I}^k(T,x,y)$ is given by \eqref{eqn:L_I_k_T_x_y} and $f_{d_i,d_j}(x,y)$ is given by~\eqref{eqn:f_i_j_x_y}.
\end{Lem}
\begin{IEEEproof}
Please refer to Appendix G.
\end{IEEEproof}

Note that $R_{M,K_n,m}$ represents an upper bound on $\Pr\left[\text{SIR}_n^{\text{pzf}}>\tau|\ell_{0,n}=m\right]$, where $m\in\{1,\cdots,M-K_n+L\}$. 
Similarly, for $M\geq 2$, the impact of the physical layer parameters $\alpha$, $\tau$ and $\lambda_h$ (captured by $R_{M,K,m}$) and the impact of the caching distribution $\mathbf T$ on $q^{\text{pzf},u}_M(\mathbf K,\mathbf T)$ are separated.
Moreover, by exploring structural properties of $q^{\text{pzf},u}_M(\mathbf K,\mathbf T)$, we have the following result.
\begin{Lem}[Properties of Upper Bound]\label{Lem:PZF_properties_UB}
The upper bound $q^{\text{pzf},u}_M(\mathbf K,\mathbf T)$ is a concave increasing function of the caching distribution $\mathbf T$. 
\end{Lem}
\begin{IEEEproof}
Please refer to Appendix H.
\end{IEEEproof}

Fig.~\ref{fig:PZF_exact_bounds} plots $q^{\text{pzf},u}_M(\mathbf K,\mathbf T)$ and $q^{\text{pzf}}_M(\mathbf K,\mathbf T)$ versus $\tau$ at different $M$ and $\mathbf K$. From Fig.~\ref{fig:PZF_exact_bounds}, we can see that $q^{\text{pzf},u}_M(\mathbf K,\mathbf T)$ bounds $q^{\text{pzf}}_M(\mathbf K,\mathbf T)$ from above and tightly matches  $q^{\text{pzf}}_M(\mathbf K,\mathbf T)$. In addition, the gap betweeen $q^{\text{pzf},u}_M(\mathbf K,\mathbf T)$ and $q^{\text{pzf}}_M(\mathbf K,\mathbf T)$ decreases as $\tau$ increases. 
Thus, $q^{\text{pzf},u}_M(\mathbf K,\mathbf T)$ can serve as a good approximation for $q^{\text{pzf}}_M(\mathbf K,\mathbf T)$. 

\subsection{Performance Optimization for PZF Receiver}



We would like to optimize $\mathbf K$ and $\mathbf T$ to maximize the STP with the PZF receiver. 
Recall that the tractable upper bound $q^{\text{pzf},u}_M(\mathbf K,\mathbf T)$
provides a good approximation for $q^{\text{pzf}}_M(\mathbf K,\mathbf T)$, as shown in Fig.~\ref{fig:PZF_exact_bounds}. In addition, $q^{\text{pzf},u}_M(\mathbf K,\mathbf T)$ 
has a much simpler form than $q^{\text{pzf}}_M(\mathbf K,\mathbf T)$.
Thus, in the following, we maximize $q^{\text{pzf},u}_M(\mathbf K,\mathbf T)$
instead of directly maximizing $q^{\text{pzf}}_M(\mathbf K,\mathbf T)$.

\begin{Prob}[Performance Optimization for PZF Receiver]\label{Prob:PZF_ub}
\small{\begin{align}
\max_{\mathbf K,\mathbf T}& \quad q_M^{\text{pzf},u}(\mathbf K,\mathbf T) \notag\\
s.t.& \quad \eqref{eqn:cache_interval_constr}, \eqref{eqn:cache_size_constr}, \eqref{eqn:DoF_allocation_PZF}.\notag
\end{align}}\normalsize
Let $(\mathbf K^*,\mathbf T^*)$ denote the optimal solution.
\end{Prob}

Since $q^{\text{pzf},u}_{M,n}(K_n,T_n)$ is an increasing function of $T_n$, by contradiction, we can easily show that files of higher popularity get more storage resources.
\begin{Lem}[Property of Optimal Solution]\label{Lem:property_solution_PZF}
The optimal solution $(\mathbf K^*,\mathbf T^*)$ of Problem~\ref{Prob:PZF_ub} satisfies $T_1^*\geq\cdots\geq T_n^*$.
\end{Lem}

When $M=1$, Problem~\ref{Prob:MRT_ub} is a convex optimization problem, and its closed-form optimal solution is given in Theorem~$4$ of our previous work \cite{Cui16TWC}. When $M\geq 2$, Problem~\ref{Prob:PZF_ub} is a mixed discrete-continuous problem with two main challenges. One is the choice of the number of DoF allocated to boost the signal power, i.e., $\mathbf K$ (discrete variables), and the other is the choice of the caching distribution of the random caching scheme, i.e., $\mathbf T$ (continuous variables).
We thus propose an equivalent alternative formulation of Problem~\ref{Prob:PZF_ub} which naturally subdivides Problem~\ref{Prob:PZF_ub} according to these two aspects.
\begin{Prob}[Equivalent Problem of Problem~\ref{Prob:PZF_ub}]\label{Prob:PZF_ub_equivalent}
\small{\begin{align}
q_M^{\text{pzf},*}= \max_{\mathbf K}& \quad q_M^{\text{pzf},*}(\mathbf K)\label{Prob:PZF_discrete}\\
s.t.& \quad \eqref{eqn:DoF_allocation_PZF}, \notag
\end{align}}\normalsize
where 
$q_M^{\text{pzf},*}(\mathbf K)$ is given by
\small{\begin{align}
q_M^{\text{pzf},*}(\mathbf K) \triangleq \max_{\mathbf T}& \quad q_M^{\text{pzf},u}(\mathbf K,\mathbf T) \label{Prob:PZF_continuous}\\
s.t.& \quad \eqref{eqn:cache_interval_constr}, \eqref{eqn:cache_size_constr}. \notag
\end{align}}\normalsize
The optimal solution to the optimization in \eqref{Prob:PZF_discrete} is $\mathbf K^*$ and the optimal solution to the optimization in~\eqref{Prob:PZF_continuous} is denoted as $\mathbf T^*(\mathbf K)$. The optimal solution to Problem~\ref{Prob:PZF_ub_equivalent} is given by $(\mathbf K^*,\mathbf T^*(\mathbf K^*))$. Note that $\mathbf T^*(\mathbf K^*)=\mathbf T^*$.
\end{Prob}



To solve Problem~\ref{Prob:PZF_ub_equivalent}, we need to search $\mathbf K$ among $M^N$ possible choices of the optimization in~\eqref{Prob:PZF_discrete}. For each possible choice of $\mathbf K$, we need to obtain $\mathbf T^*(\mathbf K)$ by solving the convex optimization problem in~\eqref{Prob:PZF_continuous}. Thus, the brute-force optimal solution to Problem~\ref{Prob:PZF_ub_equivalent} using exhaustive search is not acceptable when $N$ and $M$ are large.

In the following, we obtain a low-complexity solution to Problem~\ref{Prob:PZF_ub_equivalent} (Problem~\ref{Prob:PZF_ub}) with superior performance by an alternating optimization approach. 
In the alternating optimization approach, the parameters $\mathbf K$ and $\mathbf T$ are optimized in turn while fixing the other parameter, and the procedure is repeated iteratively until $q_M^{\text{pzf},u}(\mathbf K,\mathbf T)$ cannot be improved.
Specifically, at the $t$-th iteration, 
for given $\mathbf K(t)$, we obtain the optimal solution $\mathbf T(t)=\mathbf T^*(\mathbf K(t))$ to the convex optimization in~\eqref{Prob:PZF_continuous}, by any off-the-shelf interior-point solver (e.g., CVX~\cite{cvx}).
Then, for given $\mathbf T(t)$,  we obtain the optimal solution $\mathbf K(t+1)$ of the following discrete problem:
\small{\begin{align}
\max_{\mathbf K}& \quad q_M^{\text{pzf},u}(\mathbf K,\mathbf T(t))\label{Prob:PZF_discrete_alter}\\
s.t.& \quad \eqref{eqn:DoF_allocation_PZF}. \notag
\end{align}}\normalsize
Note that $q_{M,n}^{\text{pzf},u}(K_n,T_n)$ is a function of $K_n$ and is not affected by any $K_{n'}$, $n'\neq n$. Therefore, the discrete optimization in~\eqref{Prob:PZF_discrete_alter} can be decoupled into $N$ discrete subproblems, i.e., 
\small{\begin{align}
\max_{K_n\in\mathcal M}\quad q_{M,n}^{\text{pzf},u}(K_n,T_n(t)),\quad n\in\mathcal N.\label{Prob:PZF_dis_subproblem}
\end{align}}\normalsize
Let $K_n(t+1)$ denote the optimal solution to the discrete optimization in \eqref{Prob:PZF_dis_subproblem}. Note that $\mathbf K(t+1)=(K_n(t+1))_{n\in\mathcal N}$.
The complexity for solving the $N$ separate subproblems in~\eqref{Prob:PZF_dis_subproblem} is $O(MN)$. 
The details of the alternating optimization approach are summarized in Algorithm~\ref{alg:PZF_opt}. 
Note that Algorithm~\ref{alg:PZF_opt} stops in a finite number of iterations (which is smaller than $M^N$) due to the increase of the STP at each iteration of the alternating optimization approach. Let $(\mathbf K^{\dagger},\mathbf T^{\dagger})$ denote the near optimal solution of Problem~\ref{Prob:PZF_ub_equivalent} obtained by Algorithm~\ref{alg:PZF_opt}.

\vspace{-2mm}
\begin{algorithm} \caption{Near Optimal Solution to Problem~\ref{Prob:PZF_ub_equivalent} (Problem~\ref{Prob:PZF_ub})}
\small{\begin{algorithmic}[1]
\STATE Initialize $t=1$ and $\mathbf K(1)$ where $K_n(1)=M-1$ for all $n\in\mathcal N$. 
\STATE \textbf {repeat}
\STATE Obtain $\mathbf T(t)=\mathbf T^*(\mathbf K(t))$ by solving~\eqref{Prob:PZF_continuous} with an interior-point method.
\STATE Obtain $K_n(t+1)=\arg\max\limits_{K_n\in\mathcal M}q_{M,n}^{\text{pzf},u}(K_n,T_n(t))$ for all $n\in\mathcal N$ by solving~\eqref{Prob:PZF_dis_subproblem}.
\STATE Set $t=t+1$.
\STATE \textbf{until} $\mathbf K(t)= \mathbf K(t-1)$
\end{algorithmic}}\normalsize\label{alg:PZF_opt}
\end{algorithm}
\vspace{-4mm}

\begin{figure}[t]
\begin{center}
 \includegraphics[width=5.5cm]{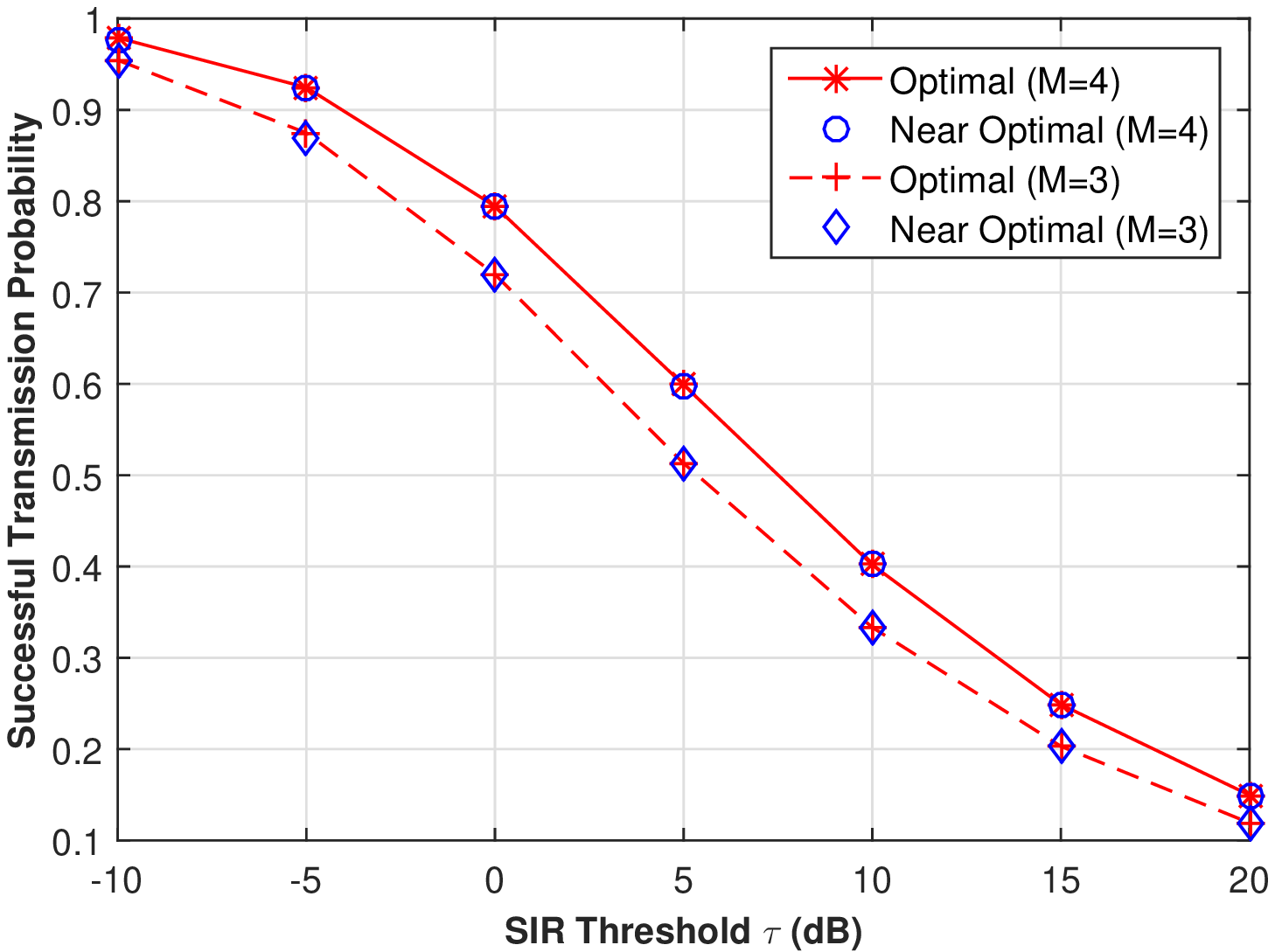}
  \end{center}
  \vspace{-4mm}
  \caption{\small{STP with the PZF receiver versus SIR threshold $\tau$.
  $N=5$, $C=3$, $\alpha=4$, $\lambda=10^{-3}$, $M=4$, $L=3$, and $a_n=\frac{n^{-\gamma}}{\sum_{n\in \mathcal N}n^{-\gamma}}$ with $\gamma=1$.}}
  \vspace{-4mm}
\label{fig:PZF_opt_vs_near_opt}
\end{figure}

We now use a numerical example to compare the optimal solution obtained by exhaustive search and the proposed near optimal solution obtained by Algorithm~\ref{alg:PZF_opt} in both STP and computation complexity. Fig.~\ref{fig:PZF_opt_vs_near_opt} plots the STP versus the SIR threshold $\tau$ at different $M$. We can see that the STP of the proposed near optimal solution nearly coincide with that of the optimal solution. In addition, when $M=4$, the numbers of convex problems we need to solve for obtaining the optimal solution and the near optimal solution are $4^5=1024$ and at most $2$, respectively. These demonstrate the applicability and effectiveness of the near optimal solution. 
In addition, Fig.~\ref{fig:opt_MRT_ub_Tn_vs_n} shows $\mathbf T^{\dagger}$ obtained by Algorithm~\ref{alg:PZF_opt} at different $M$. From Fig.~\ref{fig:opt_MRT_ub_Tn_vs_n}, we can see that files of higher popularity get more storage resources, and $T_n^{\dagger}$ decreases with $n$ more slowly when $M$ is larger, i.e., $\mathbf T^{\dagger}$ becomes more flat when $M$ is larger.

\section{Numerical Results}

In this section, we compare the proposed random caching designs with the PZF and MRC receivers with 
three baseline schemes~\cite{EURASIP15Debbah,Bharath16TCOM,Hassan15}. Baseline~$1$ (most popular) adopts the caching design in which each helper selects the $C$ most popular files to store, i.e., $T_n=1$ for $n\in\{1,\cdots,C\}$ and $T_n=0$ for $n\in\{C+1,\cdots,N\}$~\cite{EURASIP15Debbah}. Baseline~$2$ (i.i.d. file popularity) adopts the caching design in which each helper randomly selects $C$ files to store in an i.i.d. manner with file $n$ being selected with probability $a_n$~\cite{Bharath16TCOM}. Baseline 3 (uniform dist.) adopts the caching design in which each helper randomly selects $C$ different files to store, according to the uniform distribution, i.e., $T_n=\frac{C}{N}$ for all $n\in\mathcal N$~\cite{Hassan15}. The three baseline schemes all consider the MRC receiver as in the proposed random caching design with the MRC receiver. In the simulation, the popularity follows the Zipf distribution, i.e.,  $a_n=\frac{n^{-\gamma}}{\sum_{n\in \mathcal N}n^{-\gamma}}$, where $\gamma$ is the Zipf exponent. We choose $N=100$, $\tau=1$ and $\lambda=10^{-3}$.

\begin{figure}[t]
\begin{center}
\subfigure[\small{Number of receive antennas at $C=30$, $\alpha=4$ and $\gamma=0.6$.}]
{\resizebox{5.5cm}{!}{\includegraphics{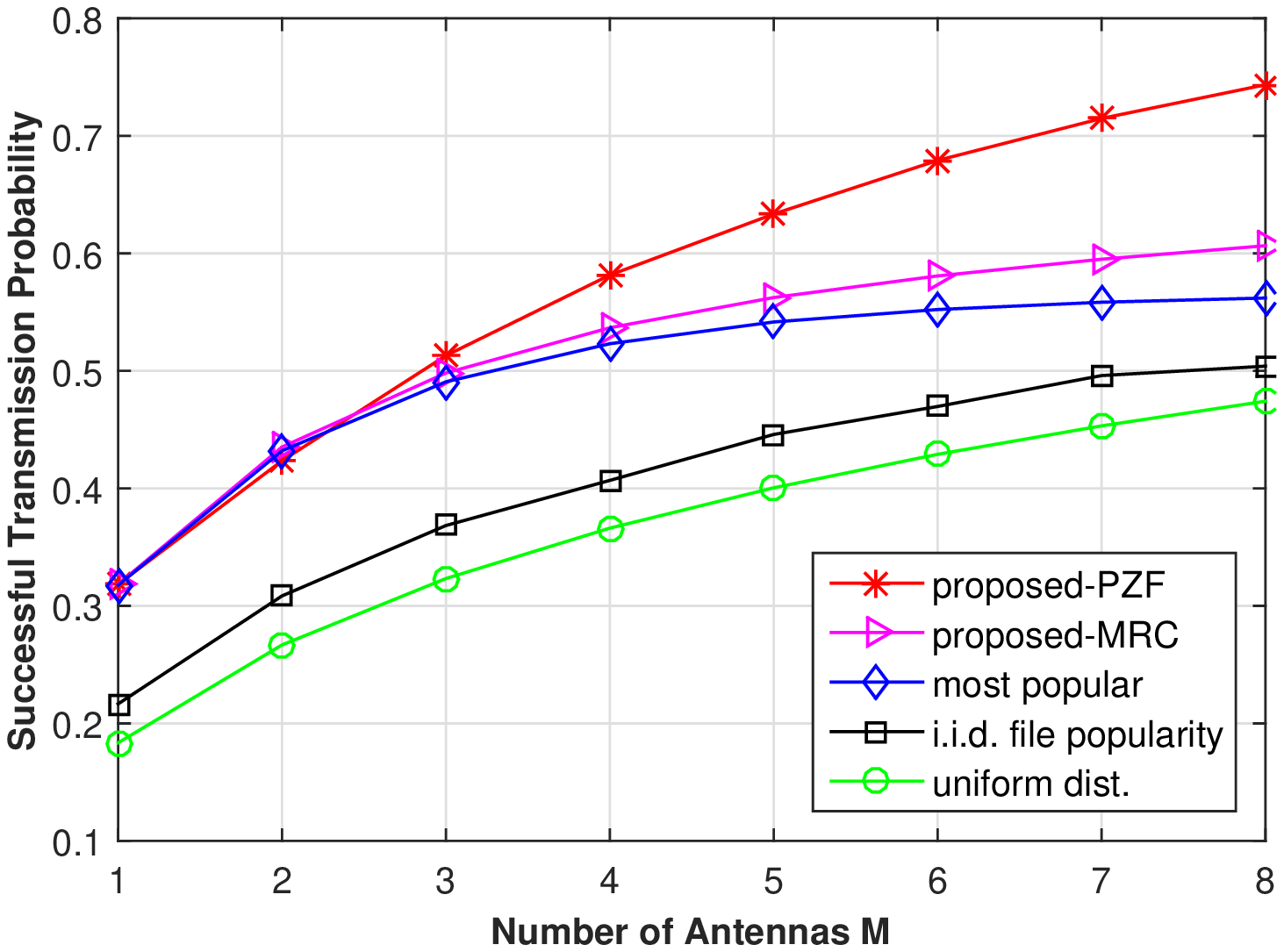}}}\quad
\subfigure[\small{Cache size at $M=6$, $\alpha=4$ and $\gamma=0.6$.}]
{\resizebox{5.5cm}{!}{\includegraphics{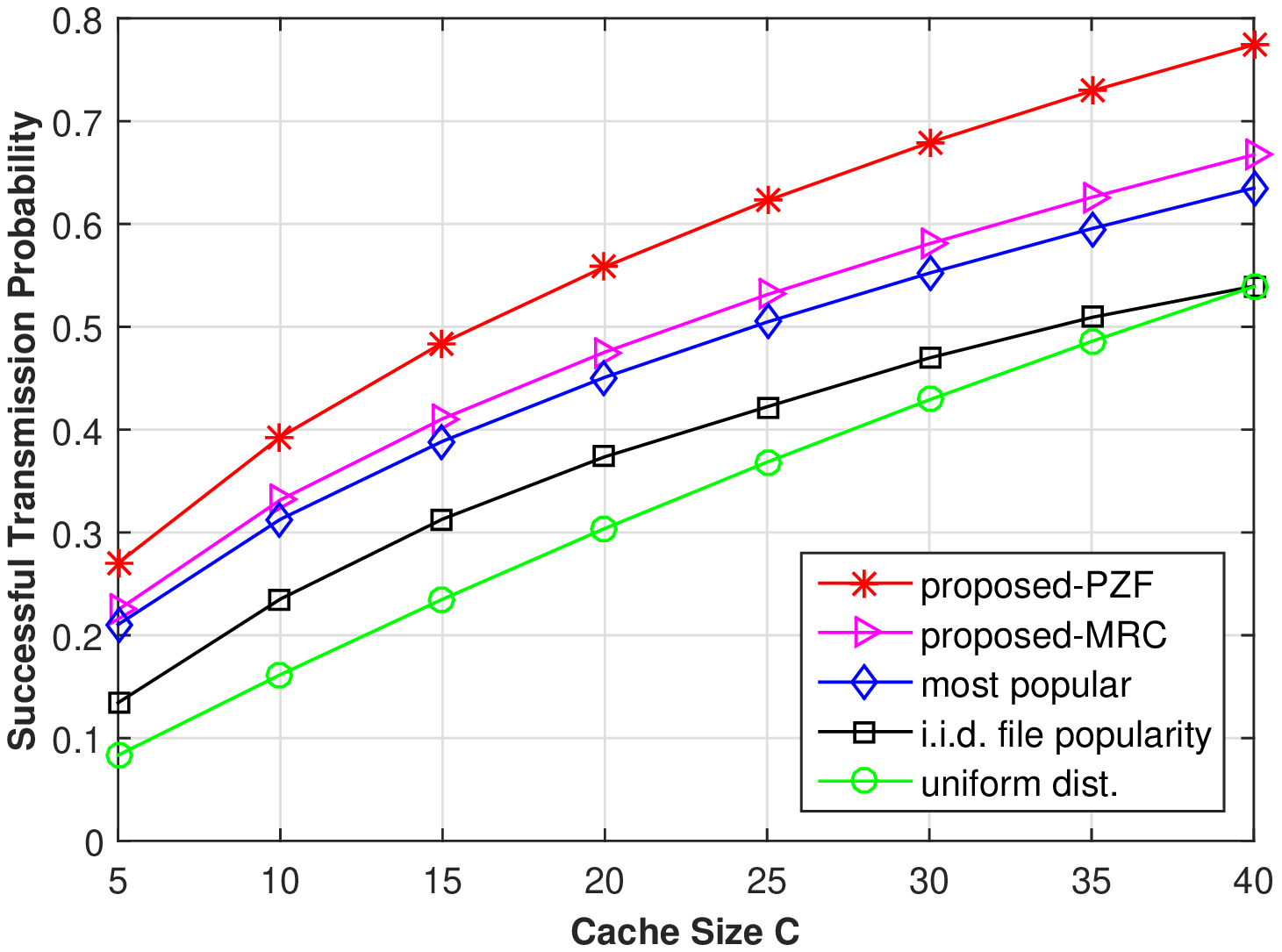}}}\quad
\subfigure[\small{Pathloss Exponent at $M=6$, $C=30$ and $\gamma=0.6$.}]
{\resizebox{5.5cm}{!}{\includegraphics{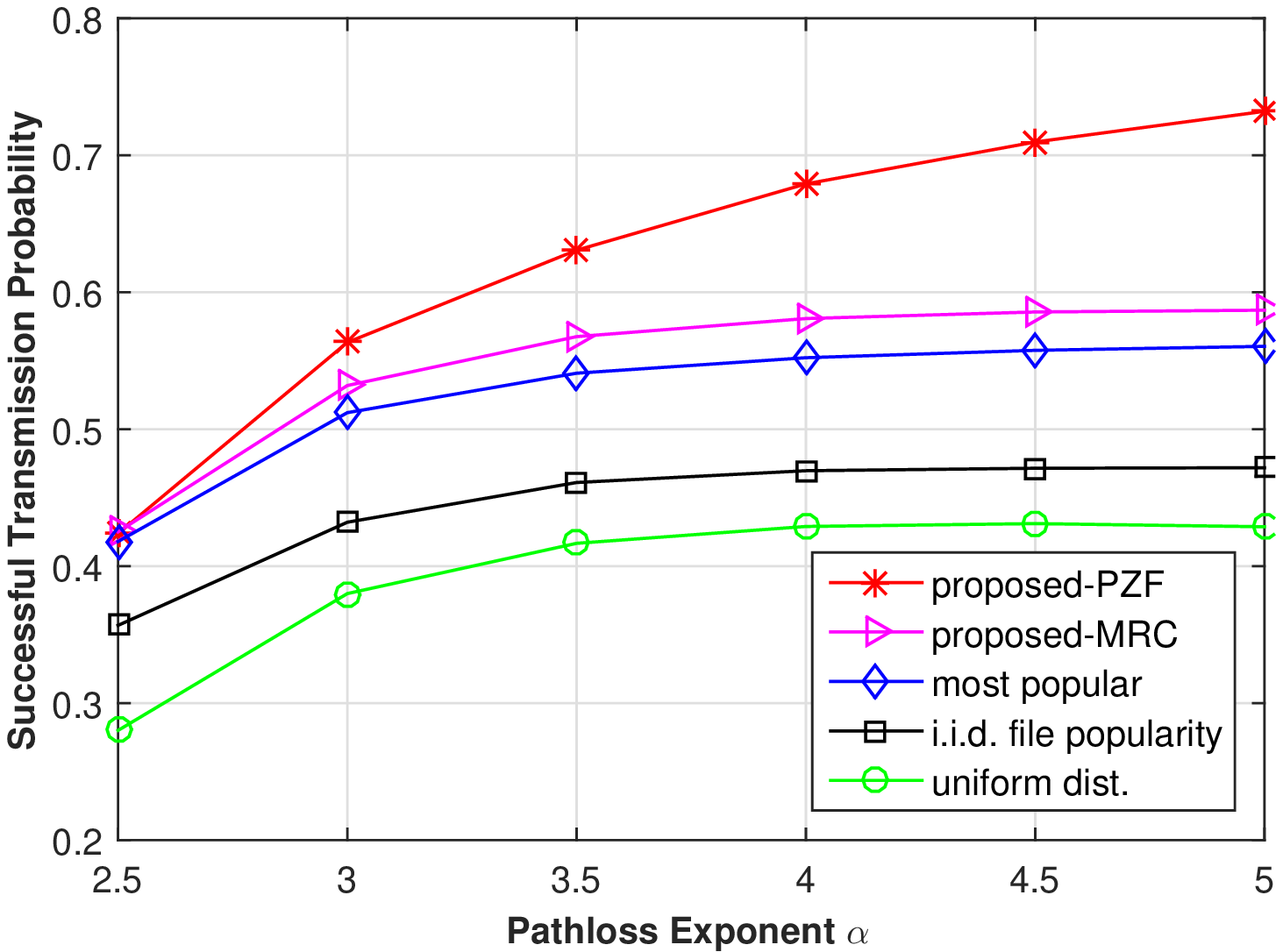}}}\quad
\subfigure[\small{Zipf Exponent at $M=6$ $C=30$ and $\alpha=4$.}]
{\resizebox{5.5cm}{!}{\includegraphics{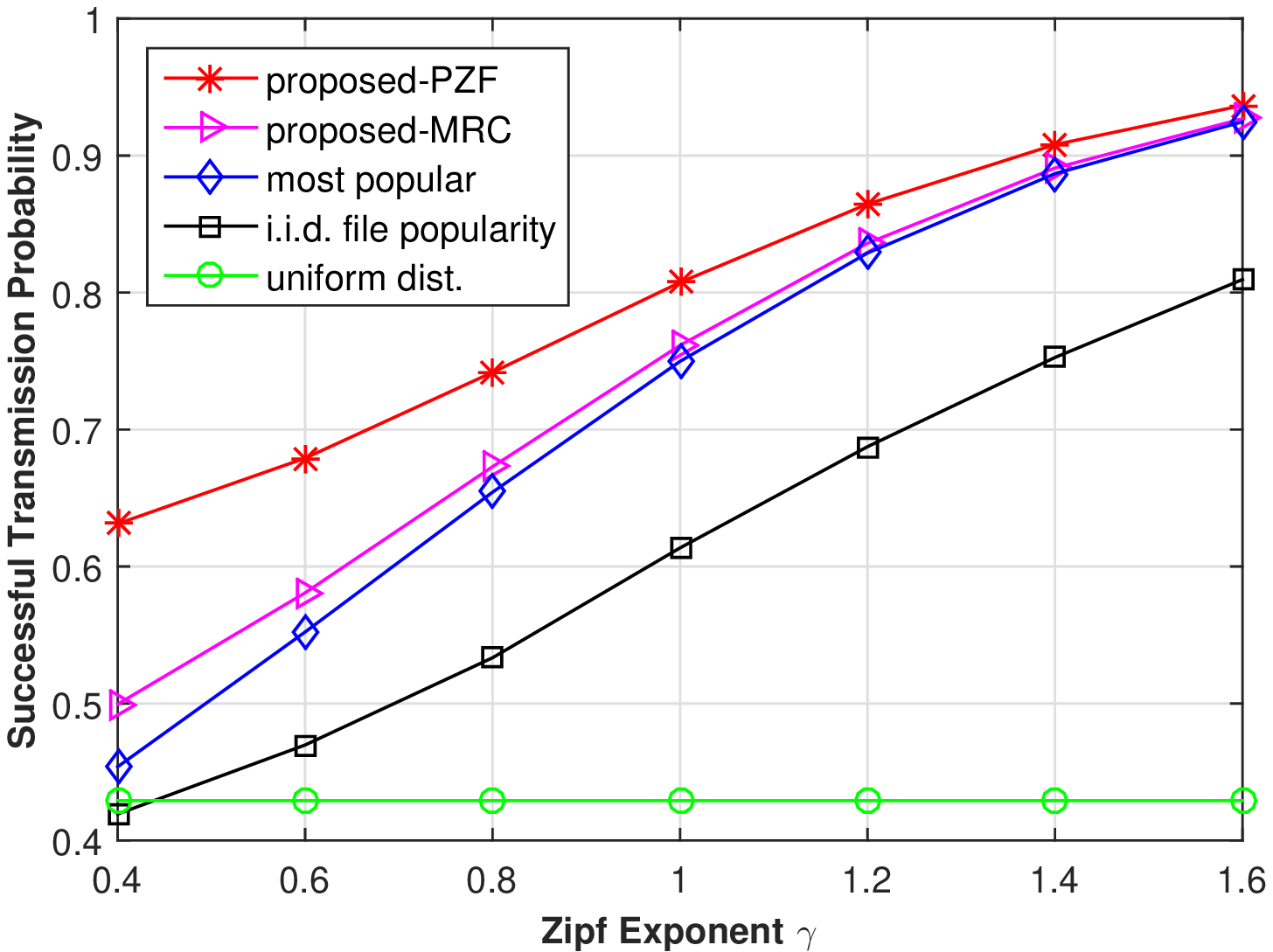}}}
\end{center}
\vspace{-4mm}
\caption{\small{STP versus number of receive antennas $M$, cache size $C$, pathloss exponent $\alpha$ and Zipf exponent $\gamma$.}}
\vspace{-6mm}
\label{fig:performance_comparison}
\end{figure}

Fig.~\ref{fig:performance_comparison} plots the STP versus $M$, $C$, $\alpha$ and $\gamma$. We can observe that the proposed designs with the MRC and PZF receivers outperform all the three baseline schemes. This is because the proposed designs can wisely exploit the storage resource.
In addition, the proposed design with the PZF receiver outperforms the proposed design with the MRC receiver. 
This is due to the fact that the proposed design with the PZF receiver can also wisely exploit antenna resource besides storage resource.
The STP gap between the two proposed designs is relatively large at large number of receive antennas $M$, large cache size $C$, large pathloss exponent $\alpha$ and small Zipf exponent $\gamma$. This demonstrates the significant benefit of making good use of antenna resource in these regions.

Specifically, 
from Fig.~\ref{fig:performance_comparison}~(a), we can see that the STP of each of the five schemes increases with $M$. This is because the receive signal power increases with $M$ for the MRC receiver, and the receive signal power increases or the interference power decreases with $M$ for the PZF receiver.
From Fig.~\ref{fig:performance_comparison}~(b), we can see that the STP of each of the five schemes increases with $C$. This is due to the fact that as $C$ increases, each helper can store more files, and the average distance between a user and its serving helper decreases.
From Fig.~\ref{fig:performance_comparison}~(c), we can see that the STP of each of the five schemes increases with $\alpha$. This is because as $\alpha$ increases, the attenuation of the receive signal power is relatively smaller than that of the interference power. 
From Fig.~\ref{fig:performance_comparison}~(d), we can see that the STP of each scheme except Baseline $3$ increases with $\gamma$. This is because as $\gamma$ increases, the tail of the popularity distribution becomes small, and a randomly requested file by a user is stored at a nearby helper with a higher probability under popularity-aware caching designs. 

\section{Conclusion}

In this paper, we considered random caching at helpers and MRC and PZF receivers at users in a large-scale cache-enabled SIMO network. First, we analyzed the STP. For each receiver, we derived a tractable expression and a tight upper bound for the STP.
The analytical results showed that for each receiver, the STP increases with the number of receive antennas $M$.  
Then, for each receiver, we considered the STP maximization via maximizing the tight upper bound on the STP. In the case of the MRC receiver, the optimization problem is non-convex, and we obtained a stationary point by solving an equivalent DC programming problem using CCCP.
In the case of the PZF receiver, the optimization problem is a mixed discrete-continuous problem, and we  obtained a low-complexity near optimal solution by an alternating optimization approach.
The optimization results indicated that files of higher popularity get more storage resources, and the optimized caching distribution becomes more flat when $M$ is larger. 

\section*{Appendix A: Proof of Theorem~\ref{Thm:STP_MRC_exact}}


First, we calculate the STP of file $n$:
\small{\begin{align}
q^{\text{mrc}}_{M,n}(T_n)=\Pr\left[\text{SIR}^{\text{mrc}}_n\geq\tau\right]
=\int_0^{\infty}\Pr\left[\text{SIR}_n^{\text{mrc}}\geq\tau|d_{\ell_{0,n}}=x\right]f_{d_{\ell_{0,n}}}(x){\rm d}x, \label{eqn:STP_file_n}
\end{align}}\normalsize
where $f_{d_{\ell_{0,n}}}(x)$ denotes the probability density function (pdf) of random variable $d_{\ell_{0,n}}$. 
Note that we have $f_{d_{\ell_{0,n}}}(x)=2\pi\lambda_h T_nx\exp\left(-\pi \lambda_h T_n x^2\right)$, as the helpers storing file $n$ form a homogeneous PPP with density $\lambda_hT_n$. To calculate $q^{\text{mrc}}_{M,n}(T_n)$, it remains to calculate $\Pr\left[\text{SIR}_n^{\text{mrc}}\geq\tau|d_{\ell_{0,n}}=x\right]$.
We rewrite 
$\text{SIR}_n^{\text{mrc}}=\frac{H^{\text{mrc}}_{\ell_{0,n},0}d_{\ell_{0,n}}^{-\alpha}}{I_n+I_{-n}}$, where $I_n\triangleq \sum_{i\in\Phi_{h,n}\setminus\{\ell_{0,n}\}}H^{\text{mrc}}_{i,0}d_i^{-\alpha}$ and $I_{-n}\triangleq \sum_{i\in\Phi_{h,-n}}H^{\text{mrc}}_{i,0}d_i^{-\alpha}$. Here,
$\Phi_{h,n}$ denotes the point process generated by helpers storing file $n$ and $\Phi_{h,-n}\triangleq \Phi_h\setminus\Phi_{h,n}$. Due to independent thinning, point processes $\Phi_{h,n}$ and $\Phi_{h,-n}$ are two independent homogeneous PPPs with density $\lambda_hT_n$ and $\lambda_h(1-T_n)$, respectively. Therefore, we have:
\small{\begin{align}
&\Pr\left[\text{SIR}_n^{\text{mrc}}\geq\tau|d_{\ell_{0,n}}=x\right]
= \Pr\left[H^{\text{mrc}}_{\ell_{0,n},0}\geq s(I_n+I_{-n})\right]\eqla \mathbb E_{I_{n}, I_{-n}}\left[\exp\left(-s(I_n+I_{-n})\right)\sum_{m=0}^{M-1}\frac{s^m(I_n+I_{-n})^m}{m!}\right]\notag
\end{align}
\begin{align}
&=\sum_{m=0}^{M-1}\frac{s^m}{m!}\sum_{k=0}^m\binom{m}{k}\mathbb E_{I_{n}}\left[I_n^k\exp\left(-sI_n\right)\right]\mathbb E_{I_{-n}}\left[I_{-n}^{m-k}\exp\left(-sI_{-n}\right)\right]
=\sum_{m=0}^{M-1}\frac{(-s)^m}{m!}\sum_{k=0}^m\binom{m}{k}\mathcal L^{(k)}_{I_n}(s)\mathcal L^{(m-k)}_{I_{-n}}(s),\label{eqn:cond_STP_file_n}
\end{align}}\normalsize
where $s=\tau x^{\alpha}$, (a) is due to $H^{\text{mrc}}_{\ell_{0,n},0}\dis \text{Gamma}(M,1)$, and $\mathcal L^{(k)}_{I}(s)\triangleq \mathbb E_{I}\left[(-I)^k\exp\left(-sI\right)\right]$ denotes the $n$th-order derivative of the Laplace transform of random variable $I$, i.e., $\mathcal L_{I}(s)\triangleq \mathbb E_{I}\left[\exp(-sI)\right]$.

Then, we calculate  $\mathcal L^{(k)}_{I_n}(s)$ and $\mathcal L^{(k)}_{I_{-n}}(s)$, respectively. $\mathcal L_{I_n}(s)$ can be calculated as follows.
\small{\begin{align}
&\mathcal L_{I_n}(s)=\mathbb E_{\Phi_{h,n},\{H^{\text{mrc}}_{i,0}\}}\left[\exp\left(-s\sum_{i\in\Phi_{h,n}\setminus\{\ell_{0,n}\}}H^{\text{mrc}}_{i,0}d_i^{-\alpha}\right)\right]
=\mathbb E_{\Phi_{h,n}}\left[\prod_{i\in\Phi_{h,n}\setminus\{\ell_{0,n}\}}\mathbb E_{H^{\text{mrc}}_{i,0}}\left[\exp\left(-sH^{\text{mrc}}_{i,0}d_i^{-\alpha}\right)\right]\right]\notag\\
&\eqlb \mathbb E_{\Phi_{h,n}}\left[\prod_{i\in\Phi_{h,n}\setminus\{\ell_{0,n}\}}\frac{1}{1+sd_i^{-\alpha}}\right]
\eqlc\exp\left(-2\pi\lambda_h T_n \int_{x}^{\infty}\left(1-\frac{1}{1+sr^{-\alpha}}\right)r{\rm d}r\right)\notag\\
&\eqld \exp\left(-\frac{2\pi\lambda_h T_n s^{\frac{2}{\alpha}}}{\alpha}B'\left(\frac{2}{\alpha},1-\frac{2}{\alpha},\frac{1}{1+\tau}\right)\right),\label{eqn:L_I_n}
\end{align}}\normalsize
where (b) is due to $H^{\text{mrc}}_{i,0}\dis \text{Exp}(1)$, (c) is obtained by using the probability generating functional of a PPP, (d) is obtained by first replacing $s^{-\frac{1}{\alpha}}r$ with $t$, and then replacing $\frac{1}{1+t^{-\alpha}}$ with $w$.
Similar to $\mathcal L_{I_n}(s)$, we have
\small{\begin{align}
&\mathcal L_{I_{-n}}(s)
=\exp\left(-\frac{2\pi\lambda_h(1-T_n)s^{\frac{2}{\alpha}}}{\alpha}B\left(\frac{2}
{\alpha},1-\frac{2}{\alpha}\right)\right).\label{eqn:L_I_m}
\end{align}}\normalsize
Based on \eqref{eqn:L_I_n} and utilizing Fa${\rm \grave{a}}$ di Bruno's formula, $\mathcal L^{(k)}_{I_n}(s)$ can be calculated as follows.
\small{\begin{align}
\mathcal L^{(k)}_{I_n}(s)&=\mathcal L_{I_{n}}(s)\sum_{(b_j)_{j=1}^k\in\mathcal M_k}\frac{k!}{b_1!b_2!\cdots b_k!}\prod_{j=1}^k\left(\frac{2\pi\lambda_h T_n}{j!}\int_{x}^{\infty}\frac{{\rm d}^j}{{\rm d}s^j}\left(\frac{1}{1+sr^{-\alpha}}\right)r{\rm d}r\right)^{b_j}\notag\\
&\eqle(-1)^k\mathcal L_{I_{n}}(s)\sum_{(b_j)_{j=1}^k\in\mathcal M_k}\frac{k!}{b_1!b_2!\cdots b_k!}\prod_{j=1}^k\left(2\pi\lambda_h T_n\int_{x}^{\infty}\frac{r^{-j\alpha+1}}{(1+sr^{-\alpha})^{j+1}}{\rm d}r\right)^{b_j}\notag\\
&\eqlf\left(-\frac{1}{s}\right)^k\mathcal L_{I_{n}}(s)\sum_{(b_j)_{j=1}^k\in\mathcal M_k}\frac{k!}{b_1!b_2!\cdots b_k!}\prod_{j=1}^k\left(\frac{2\pi\lambda_h T_n}{\alpha}s^{\frac{2}{\alpha}}B'\left(\frac{2}{\alpha}+1,j-\frac{2}{\alpha},\frac{1}{1+\tau}\right)\right)^{b_j},\label{eqn:L_k_I_n}
\end{align}}\normalsize
where (e) is due to $\frac{{\rm d}^{j}}{{\rm d}s^{j}}\frac{1}{1+sr^{-\alpha}}=  \frac{(-1)^j (j!)r^{-j\alpha}}{\left(1+sr^{-\alpha}\right)^{j+1}}$ and $\prod_{j=1}^{k}(-1)^{jb_j}=(-1)^{\sum_{j=1}^{k}jb_j}=(-1)^{k}$, and (f) is obtained by first replacing $s^{-\frac{1}{\alpha}}r$ with $t$ and then replacing $\frac{1}{1+t^{-\alpha}}$ with $w$. Similar to $\mathcal L^{(k)}_{I_n}(s)$, we have
\small{\begin{align}
\mathcal L^{(k)}_{I_{-n}}(s)
&=\left(-\frac{1}{s}\right)^k\mathcal L_{I_{-n}}(s)\sum_{(b_j)_{j=1}^k\in\mathcal M_k}\frac{k!}{b_1!b_2!\cdots b_k!}\prod_{j=1}^k\left(\frac{2\pi\lambda_h (1-T_n)}{\alpha}s^{\frac{2}{\alpha}}B\left(\frac{2}{\alpha}+1,j-\frac{2}{\alpha}\right)\right)^{b_j}.\label{eqn:L_k_I_m}
\end{align}}\normalsize

Finally, substituting \eqref{eqn:L_k_I_n} and \eqref{eqn:L_k_I_m} into \eqref{eqn:cond_STP_file_n}, we have
\small{\begin{align}
\Pr\left[\text{SIR}_n^{\text{mrc}}\geq\tau|d_{\ell_{0,n}}=x\right]=\sum_{m=0}^{M-1}\frac{1}{m!}\sum_{k=0}^m\binom{m}{k}\tilde{\mathcal L}^{(k)}_{I_n}(s)\tilde{\mathcal L}^{(m-k)}_{I_{-n}}(s),\label{eqn:cond_STP_file_n_exact}
\end{align}}\normalsize
where $\tilde{\mathcal L}_{I_n}^{(k)}(s)\triangleq {\mathcal L^{(k)}_{I_n}(s)}/{\left(-\frac{1}{s}\right)^k}$ and $\tilde{\mathcal L}_{I_{-n}}^{(k)}(s)\triangleq {\mathcal L^{(k)}_{I_{-n}}(s)}/{\left(-\frac{1}{s}\right)^k}$. Substituting \eqref{eqn:cond_STP_file_n_exact} into \eqref{eqn:STP_file_n}, we can obtain $q^{\text{mrc}}_{M,n}(T_n)$.
According to total probability theorem, we complete the proof of Theorem~\ref{Thm:STP_MRC_exact}.

\vspace{-4mm}
\section*{Appendix B: Proof of Lemma~\ref{Lem:bounds_MRC}}

We obtain an upper bound on the STP of file $n$ as follows.
\small{\begin{align}
q^{\text{mrc}}_{M,n}(T_n)&=1-\int_0^{\infty}\Pr\left[\text{SIR}_n^{\text{mrc}}<\tau|d_{\ell_{0,n}}=x\right]f_{d_{\ell_{0,n}}}(x){\rm d}x
\eqla 1-\int_0^{\infty}\mathbb E_{I_n,I_{-n}}\left[\frac{\gamma(M,s(I_n+I_{-n}))}{\Gamma(M)}\right]f_{d_{\ell_{0,n}}}(x){\rm d}x\notag\\
&\stackrel{(b)}{<}1-\int_0^{\infty}E_{I_n,I_{-n}}\left[(1-\exp\left(-S_Ms(I_n+I_{-n})\right))^M\right]f_{d_{\ell_{0,n}}}(x){\rm d}x\notag\\
&=1-\int_0^{\infty}E_{I_n,I_{-n}}\left[\sum_{k=0}^M\binom{M}{k}(-1)^k\exp(-kS_Ms(I_n+I_{-n}))\right]f_{d_{\ell_{0,n}}}(x){\rm d}x\notag\\
&=1-\sum_{k=0}^M\binom{M}{k}(-1)^k\int_0^{\infty}\mathcal L_{I_n}(kS_Ms)\mathcal L_{I_{-n}}(kS_Ms)f_{d_{\ell_{0,n}}}(x){\rm d}x\triangleq q_{M,n}^{\text{mrc},u}(T_n),\label{eqn:MRC_UB_STP_file_n}
\end{align}}\normalsize
where (a) is due to $H^{\text{mrc}}_{\ell_{0,n},0}\dis \text{Gamma}(M,1)$ and (b) is based on a lower bound on the incomplete gamma function, i.e., $\left(1-e^{-S_ab}\right)^a<\frac{\gamma(a,b)}{\Gamma(a)}$ for $a>1$.
Note that $\mathcal L_{I_n}(kS_Ms)$ and $\mathcal L_{I_{-n}}(kS_Ms)$ are given by \eqref{eqn:L_I_n} and \eqref{eqn:L_I_m}, respectively. Substituting  \eqref{eqn:L_I_n} and \eqref{eqn:L_I_m} into \eqref{eqn:MRC_UB_STP_file_n}, we have
\small{\begin{align}
q_{M,n}^{\text{mrc},u}(T_n) &= 1-\sum_{k=0}^M\binom{M}{k}(-1)^k2\pi\lambda_hT_n\int_0^{\infty}x\exp(-(\pi\lambda_hT_n+A)x^2){\rm d}x
\eqlc 1-\sum_{k=0}^M\binom{M}{k}(-1)^k\frac{\pi\lambda_hT_n}{\pi\lambda_hT_n+A},\notag
\end{align}}\normalsize
where $A\triangleq \frac{2}{\alpha}(kS_M\tau)^{\frac{2}{\alpha}}B'(\frac{2}{\alpha},1-\frac{2}{\alpha},\frac{1}{1+kS_M\tau})
+\frac{2}{\alpha}(\frac{1}{T_n}-1)(kS_M\tau)^{\frac{2}{\alpha}}B(\frac{2}{\alpha},1-\frac{2}{\alpha})$ and (c) is due to $\int_{0}^{\infty}x\exp(-cx^2){\rm d}x=\frac{1}{2c}$. After some algebraic
manipulations, we can obtain $q_{M,n}^{\text{mrc},u}(T_n)$ and the corresponding upper bound $q_M^{\text{mrc},u}(\mathbf T)$ on $q_M^{\text{mrc}}(\mathbf T)$.

\vspace{-4mm}
\section*{Appendix C: Proof of Lemma~\ref{Lem:STP_low_SIR}}

First, we calculate the conditional outage probability of file $n$ conditioned on $d_{\ell_{0,n}}=x$. Similar to the calculation of $\Pr\left[\text{SIR}_n^{\text{mrc}}\geq\tau|d_{\ell_{0,n}}=x\right]$ in Appendix~A, we have
\small{\begin{align}
&\Pr\left[\text{SIR}_n^{\text{mrc}}<\tau|d_{\ell_{0,n}}=x\right]= \Pr\left[H^{\text{mrc}}_{\ell_{0,n},0}< s(I_n+I_{-n})\right]
\eqla \mathbb E_{I_{n}, I_{-n}}\left[\exp\left(-s(I_n+I_{-n})\right)\sum_{m=M}^{\infty}\frac{s^m(I_n+I_{-n})^m}{m!}\right]\notag\\
&=\sum_{m=M}^{\infty}\frac{(-s)^m}{m!}\sum_{k=0}^m\binom{m}{k}\mathcal L^{(k)}_{I_n}(s)\mathcal L^{(m-k)}_{I_{-n}}(s)
=\sum_{m=M}^{\infty}\frac{1}{m!}\sum_{k=0}^m\binom{m}{k}\tilde{\mathcal L}^{(k)}_{I_n}(s)\tilde{\mathcal L}^{(m-k)}_{I_{-n}}(s),\notag
\end{align}}\normalsize
where (a) is due to $H^{\text{mrc}}_{\ell_{0,n},0}\dis \text{Gamma}(M,1)$.
Then, by removing the condition of $\Pr[\text{SIR}_n^{\text{mrc}}<\tau|d_{\ell_{0,n}}=x]$ on $d_{\ell_{0,n}}=x$, we obtain the outage probability of file $n$:
\small{\begin{align}
\overline{q}^{\text{mrc}}_{M,n}(T_n)=\int_{0}^{\infty}f_{d_{\ell_{0,n}}}(x)\sum_{m=M}^{\infty}\frac{1}{m!}\sum_{k=0}^m\binom{m}{k}\tilde{\mathcal L}^{(k)}_{I_n}(s)\tilde{\mathcal L}^{(m-k)}_{I_{-n}}(s){\rm d}x.\notag
\end{align}}\normalsize

Next, we calculate the asymptotic outage probability of file $n$ in the low SIR threshold regime, i.e., $ \lim_{\tau\to 0}\overline{q}^{\text{mrc}}_{M,n}(T_n)$. According to dominated convergence theorem, we have:
\small{\begin{align}
\lim_{\tau\to 0}\overline{q}^{\text{mrc}}_{M,n}(T_n)&= \lim_{\tau\to 0}\int_{0}^{\infty}f_{d_{\ell_{0,n}}}(x)\sum_{m=M}^{\infty}\frac{1}{m!}\sum_{k=0}^m\binom{m}{k}\tilde{\mathcal L}^{(k)}_{I_n}(s)\tilde{\mathcal L}^{(m-k)}_{I_{-n}}(s){\rm d}x\notag
\end{align}
\begin{align}
&= \int_{0}^{\infty}\lim_{\tau\to 0}f_{d_{\ell_{0,n}}}(x)\sum_{m=M}^{\infty}\frac{1}{m!}\sum_{k=0}^m\binom{m}{k}\tilde{\mathcal L}^{(k)}_{I_n}(s)\tilde{\mathcal L}^{(m-k)}_{I_{-n}}(s){\rm d}x.\label{eqn:MRC_OP_n_step1}
\end{align}}\normalsize
We note that $B^{'}(a,b,z)=\frac{(1-z)^{b}}{b}+o\left((1-z)^{b}\right)$ as $z\to1$.
Then, as $\tau\to 0$, we have $B^{'}\left(\frac{2}{\alpha},1-\frac{2}{\alpha},\frac{1}{1+\tau}\right) = \frac{\left(\tau\right)^{1-\frac{2}{\alpha}}}{1-\frac{2}{\alpha}}+o\left(\tau^{1-\frac{2}{\alpha}}\right)$ and $B^{'}\left(1+\frac{2}{\alpha},a-\frac{2}{\alpha},\frac{1}{1+\tau}\right)= \frac{(\tau)^{a-\frac{2}{\alpha}}}{a-\frac{2}{\alpha}}+o\left(\tau^{a-\frac{2}{\alpha}}\right)$.
Based on these two asymptotic expressions, as $\tau\to 0$, we have:
\small{\begin{align}
\tilde{\mathcal L}^{(k)}_{I_{n}}(s)=\tau^{k}\sum_{(b_j)_{j=1}^k\in\mathcal M_k}\frac{k!}{b_1!b_2!\cdots b_k!}\prod_{j=1}^k\left(\frac{\frac{2\pi}{\alpha}\lambda_hT_nx^2}{j-\frac{2}{\alpha}}\right)^{b_j}+o(\tau^{k}).\label{eqn:L_k_I_n_asymp}
\end{align}}\normalsize
On the other hand, as $\tau\to 0$, we have
\small{\begin{align}
\tilde{\mathcal L}^{(k)}_{I_{-n}}(s)&=\mathcal L_{I_{-n}}(s)\sum_{{(b_j)_{j=1}^k\in\mathcal M_k}}\tau^{\frac{2}{\alpha}\sum_{j=1}^kb_j}\frac{k!}{b_1!b_2!\cdots b_k!}\prod_{j=1}^k\left(\frac{2\pi\lambda_h (1-T_n)x^2}{\alpha}B\left(\frac{2}{\alpha}+1,j-\frac{2}{\alpha}\right)\right)^{b_j}\notag\\
&\eqla\tau^{\frac{2}{\alpha}}k!\frac{2\pi\lambda_h (1-T_n)x^2}{\alpha}B\left(\frac{2}{\alpha}+1,k-\frac{2}{\alpha}\right)+o(\tau^{\frac{2}{\alpha}}),\label{eqn:L_k_I_m_asymp}
\end{align}}\normalsize
where (a) is due to the fact that  $b_1=\cdots=b_{k-1}=0$ and $b_{k}=1$ is the dominant term in $\sum_{{(b_j)_{j=1}^k\in\mathcal M_k}}\tau^{\frac{2}{\alpha}\sum_{j=1}^kb_j}$ as $\tau \to 0$. Substituting \eqref{eqn:L_k_I_n_asymp} and \eqref{eqn:L_k_I_m_asymp} into  \eqref{eqn:MRC_OP_n_step1}, we have
\small{\begin{align}
&\lim_{\tau\to 0}\overline{q}^{\text{mrc}}_{M,n}(T_n)= \sum_{m=M}^{\infty}\frac{1}{m!}\sum_{k=0}^m\binom{m}{k} \tau^{k+\frac{2}{\alpha}} \int_{0}^{\infty}f_{d_{\ell_{0,n}}}(x)\left(\sum_{(b_j)_{j=1}^k\in\mathcal M_k}\frac{k!}{b_1!b_2!\cdots b_k!}\prod_{j=1}^k\left(\frac{\frac{2\pi}{\alpha}\lambda_hT_nx^2}{j-\frac{2}{\alpha}}\right)^{b_j}\right)\notag\\
&\quad \times\left((m-k)!\frac{2\pi\lambda_h (1-T_n)x^2}{\alpha}B\left(\frac{2}{\alpha}+1,m-k-\frac{2}{\alpha}\right)\right){\rm d}x +o(\tau^{k+\frac{2}{\alpha}})\notag\\
&\eqlb \tau^{\frac{2}{\alpha}} \frac{2\pi\lambda_h (1-T_n)}{\alpha}\sum_{m=M}^{\infty}B\left(\frac{2}{\alpha}+1,m-\frac{2}{\alpha}\right) \int_{0}^{\infty}x^2f_{d_{\ell_{0,n}}}(x){\rm d}x +o(\tau^{\frac{2}{\alpha}})
=\overline{q}^{\text{mrc}}_{M,0,n}(T_n)+o(\tau^{\frac{2}{\alpha}}),\label{eqn:MRC_OP_n_step2}
\end{align}}\normalsize
where (b) is due to the fact that  $k=0$ is the dominant term in $\sum_{k=0}^m\tau^{k+\frac{2}{\alpha}}$ as $\tau \to 0$.

Finally, based on \eqref{eqn:MRC_OP_n_step2}, we have $\lim_{\tau\to 0}\overline{q}^{\text{mrc}}_{M}(\mathbf T)=\overline{q}^{\text{mrc}}_{M,0}(\mathbf T)+o(\tau^{\frac{2}{\alpha}})$.

\vspace{-4mm}
\section*{Appendix D: Proof of Lemma~\ref{Lem:opt_subprob}}

The Lagrange function of Problem~\ref{Prob:MRT_ub_equivalent_sub} is given by
\small{\begin{align}
L(\mathbf T,\bm \eta,\bm \lambda, v)&=R_o(\mathbf T)+\sum_{n\in\mathcal N}a_nf_e(T_n^{\dagger}(t-1))T_n-\sum_{n\in\mathcal N}\eta_nT_n+\sum_{n\in\mathcal N}\lambda_n(T_n-1)+v\left(\sum_{n\in\mathcal N}T_n-C\right),\notag
\end{align}}\normalsize
where $\bm \eta\triangleq(\eta_n)_{n\in\mathcal N}$ and $\bm \lambda\triangleq (\lambda_n)_{n\in\mathcal N}$ are the Lagrange multipliers associated with the inequality constraints $T_n\geq 0$, $n\in\mathcal N$ and $T_n\leq 1$, $n\in\mathcal N$, respectively. $v$ is the Lagrange multiplier associated with the equality constraint $\sum_{n\in\mathcal N}T_n=C$. Thus, we have
\small{\begin{align}
\frac{\partial L(\mathbf T,\bm \eta,\bm \lambda, v)}{\partial T_n}=-a_nf_o(T_n)+a_nf_e(T_n^{\dagger}(t-1))-\eta_n+\lambda_n+v.\notag
\end{align}}\normalsize
Since strong duality holds, primal optimal $\mathbf T^{\dagger}(t)$ and dual optimal $\bm \eta^*$, $\bm \lambda^*$ and $v^{\dagger}(t)$  satisfy KKT conditions: (i) primal constraints \eqref{eqn:cache_interval_constr} and \eqref{eqn:cache_size_constr}; (ii) dual constraints $\eta_n\geq0$ and $\lambda_n\geq0$ for all  $n\in\mathcal N$; (iii) complementary slackness $\eta_nT_n = 0$ and  $\lambda_n(T_n-1)=0$ for all $n\in\mathcal N$; and (iv) $\frac{\partial L(\mathbf T,\bm \eta,\bm \lambda, v)}{\partial T_n}=0$. By (i)-(iv),
we have the following results: if $v>a_n(f_o(0)
-f_e(T_n^{\dagger}(t-1)))$, then $T_n=0$; if $v<a_n(f_o(1)-f_e(T_n^{\dagger}(t-1)))$, then $T_n=1$; if $a_n(f_o(1)-f_e(T_n^{\dagger}(t-1)))<v<a_n(f_o(0)
-f_e(T_n^{\dagger}(t-1)))$, then $T_n$ satisfies $f_o(T_n)=f_e(T_n^{\dagger}(t-1))+\frac{v}{a_n}$. Combining with \eqref{eqn:cache_size_constr}, we can prove Lemma~\ref{Lem:opt_subprob}.

\vspace{-4mm}
\section*{Appendix E: Proof of Lemma~\ref{Lem:property_solution_MRC}}


We prove Lemma~\ref{Lem:property_solution_MRC} by mathematical induction. First, according to the initialization, we have $T_n^{\dagger}(1)=T_{n'}^{\dagger}(1)=\frac{C}{N}$ for all $n'>n$. 
Assume $T_n^{\dagger}(t)\geq T_{n'}^{\dagger}(t)$ for all $n'>n$, where $t\geq 1$. Next, we show $T_n^{\dagger}(t+1)\geq T_{n'}^{\dagger}(t+1)$ for all $n'>n$.
Since $a_n>a_{n'}$ and $T_n^{\dagger}(t)\geq T_{n'}^{\dagger}(t)$ for all $n'>n$, and $f_e(x)$ is a decreasing function, we have $f_e(T_n^{\dagger}(t))+\frac{v^{\dagger}(t+1)}{a_n}<f_e(T_{n'}^{\dagger}(t))+\frac{v^{\dagger}(t+1)}{a_{n'}}$. Combining with the fact that $f_o(x)$ is a decreasing function, we have (i) if $f_o(0)\leq f_e(T_{n'}^{\dagger}(t))+\frac{v^{\dagger}(t+1)}{a_{n'}}$, then $T_n^{\dagger}(t+1)\geq T_{n'}^{\dagger}(t+1)=0$;  (ii) if $f_o(1)\geq f_e(T_{n}^{\dagger}(t))+\frac{v^{\dagger}(t+1)}{a_{n}}$, then $T^{\dagger}_{n'}(t+1)\leq T^{\dagger}_n(t+1)=1$; (iii) if $f_o(1)<f_e(T_{n}^{\dagger}(t))+\frac{v^{\dagger}(t+1)}{a_{n}}<f_e(T_{n'}^{\dagger}(t))+\frac{v^{\dagger}(t+1)}{a_{n'}}<f_o(0)$, then $0<T^{\dagger}_{n'}(t+1)<T^{\dagger}_n(t+1)<1$. Therefore, we can show Lemma~\ref{Lem:property_solution_MRC}.

\vspace{-4mm}
\section*{Appendix F: Proof of Theorem~\ref{Thm:STP_PZF_exact}}

When $K_n=M$,  the PZF receiver reduces to the MRC receiver and $q_{M,n}^{\text{pzf}}(K_n,T_n)=q_{M,n}^{\text{mrc}}(T_n)$. When $K_n<M$, by the law of total probability, we have
\small{\begin{align}
q^{\text{pzf}}_{M,n}(K_n,T_n)&=\Pr\left[\ell_{0,n}>M-K_n\right]\Pr\left[\text{SIR}_n^{\text{pzf}}>\tau|\ell_{0,n}>M-K_n\right]\notag\\
&+\sum_{m=1}^{M-K_n}\Pr\left[\ell_{0,n}=m\right]\Pr\left[\text{SIR}_n^{\text{pzf}}>\tau|\ell_{0,n}=m\right],\label{eqn:PZF_STP_n_appendix}
\end{align}}\normalsize
where $\Pr[\ell_{0,n}=m]=T_n(1-T_n)^{m-1}$ and $\Pr[\ell_{0,n}>m]=(1-T_n)^{m-1}$. To calculate  $q^{\text{pzf}}_{M,n}(K_n,T_n)$, it remains to calculate $\Pr\left[\text{SIR}_n^{\text{pzf}}>\tau|\ell_{0,n}=m\right]$ and $\Pr\left[\text{SIR}_n^{\text{pzf}}>\tau|\ell_{0,n}>M-K_n\right]$.

When $\ell_{0,n}>M-K_n$, we have
\small{\begin{align}
&\Pr\left[\text{SIR}_n^{\text{pzf}}>\tau|\ell_{0,n}>M-K_n\right]\notag\\
&=\int_{0}^{\infty}\int_{0}^{x}g_{d_{\ell_{0,n}},d_{M-K_n}}(T_n,x,y)
\Pr\left[\frac{H_{\ell_{0,n},0}^{\text{pzf}}d_{\ell_{0,n}}^{-\alpha}}{I_n+I_{-n}}>\tau\Big{|}d_{\ell_{0,n}}=x,d_{M-K_n}=y\right]
{\rm d}y{\rm d}x,\label{eqn:PZF_STP_case1}
\end{align}}\normalsize
where $I_n \triangleq \sum\limits_{i\in \Phi_{h,n} \setminus\{\ell_{0,n}\}}H_{i,0}^{\text{pzf}}d_i^{-\alpha}$, $I_{-n}\triangleq\sum\limits_{i\in \Phi_{h,-n} \setminus\{1,2,\cdots,M-K_n\}}H_{i,0}^{\text{pzf}}d_i^{-\alpha}$ and $g_{d_{\ell_{0,n}},d_{M-K_n}}(T_n,x,y)$ denotes the joint pdf of $d_{\ell_{0,n}}$ and $d_{M-K_n}$ when $\ell_{0,n}>M-K_n$. Similar to the calculation of $\Pr\left[\text{SIR}_n^{\text{mrc}}\geq\tau|d_{\ell_{0,n}}=x\right]$ in \eqref{eqn:cond_STP_file_n} in Appendix~A, we have
\small{\begin{align}
\Pr\left[\frac{H_{\ell_{0,n},0}^{\text{pzf}}d_{\ell_{0,n}}^{-\alpha}}{I_n+I_{-n}}>\tau\Big{|}d_{\ell_{0,n}}=x,d_{M-K_n}=y\right] = \sum_{m=0}^{K_n-1}\frac{1}{m!}\sum_{k=0}^m\binom{m}{k}\tilde{\mathcal L}_{I}^k(T_n,x,x)\tilde{\mathcal L}_{I}^{m-k}(1-T_n,x,y).\label{eqn:PZF_cond_case1}
\end{align}}\normalsize
Now, we calculate $g_{d_{\ell_{0,n}},d_{M-K_n}}(T_n,x,y)$. Note that the pdf of the distance of the $i$-th nearest 

\noindent point in a homogeneous PPP with density $\lambda$ is $h_{d_i}(x,\lambda)=\frac{2\pi^i\lambda^ix^{2i-1}}{(i-1)!}\exp(-\pi\lambda x^2)$.
Due to the independence between $\Phi_{h,n}$ and $\Phi_{h,-n}$, we have
\small{\begin{align}
&g_{d_{\ell_{0,n}},d_{M-K_n}}(T_n,x,y)
=\frac{h_{d_1}(x,\lambda_h T_n)h_{d_{M-K_n}}(y,\lambda_h(1-T_n))}{\Pr[\ell_{0,n}>M-K_n]}.\label{eqn:g_in_appendix}
\end{align}}\normalsize

When $\ell_{0,n}=m\in\{1,2,\cdots,M-K_n\}$, we have
\small{\begin{align}
\hspace{-2mm}\Pr\left[\text{SIR}_n^{\text{pzf}}>\tau|\ell_{0,n}=m\right] = \int_{0}^{\infty}\int_{0}^{y}f_{d_m,d_{M-K_n+1}}(x,y)\Pr\left[\frac{H_{m,0}^{\text{pzf}}
d_{m}^{-\alpha}}{I_{-}}>\tau\Big{|}d_{m}=x,d_{M-K_n+1}=y\right]
{\rm d}x{\rm d}y,\label{eqn:PZF_STP_case2}
\end{align}}\normalsize
where $I_{-} \triangleq \sum_{i\in\Phi_h\setminus\{1,\cdots,M-K_n+1\}} H_{i,0}^{\text{pzf}}d_{i}^{-\alpha}$ and $f_{d_m,d_{M-K_n+1}}(x,y)$ denotes the joint pdf of $d_{m}$ and $d_{M-K_n+1}$ when $m<M-K_n+1$. Similar to the calculation of $\Pr\left[\text{SIR}_n^{\text{mrc}}\geq\tau|d_{\ell_{0,n}}=x\right]$ in \eqref{eqn:cond_STP_file_n} in Appendix~A, we have
\small{\begin{align}
\Pr\left[\frac{H_{m,0}^{\text{pzf}}
d_{m}^{-\alpha}}{I_{-}}>\tau\Big{|}d_{m}=x,d_{M-K_n+1}=y\right]
=\sum_{m=0}^{K_n-1}\frac{1}{m!}\sum_{k=0}^m\binom{m}{k}\tilde{\mathcal L}_{I}^k(T_n,x,x)\tilde{\mathcal L}_{I}^{m-k}(1-T_n,x,y).\label{eqn:PZF_cond_case2}
\end{align}}\normalsize
Now, we calculate $f_{d_m,d_{M-K_n+1}}(x,y)$. Consider four non-overlapping areas $\mathcal A_1=\mathcal B(0,d_m)$, $\mathcal A_2=\mathcal B(0,d_m+{\rm d}r_1)\setminus \mathcal B(0,d_m)$, $\mathcal A_3=\mathcal B(0,d_{M-K_n+1})\setminus \mathcal B(0,d_m+{\rm d}r_1)$ and $\mathcal A_4=\mathcal B(0,d_{M-K_n+1}+{\rm d}r_2)\setminus \mathcal B(0,d_{M-K_n+1})$, where $\mathcal B(0,d)$ denotes the ball centered at the origin with radius $d$. According to the definition of PPP, the joint probability that nodes $m$ and $M-K_n+1$ belong to $\mathcal A_2$ and $\mathcal A_4$, respectively, is given by 
\small{\begin{align}
&\Pr[\text{node $m$ in } \mathcal A_2,\ \text{node $M-K_n+1$ in } \mathcal A_4]
=\begin{cases}
P_1P_2P_3P_4 \hspace{10mm} d_m\leq d_{M-K_n+1}\\
0 \hspace{25mm} \text{otherwise}
\end{cases},\label{eqn:node-m-in-A-2}
\end{align}}\normalsize
where $P_1 = \Pr[\text{$m-1$ nodes in } \mathcal A_1]= \frac{(\pi\lambda_h d_m^2)^{m-1}}{(m-1)!}\exp(-\pi\lambda_h d_m^2)$, $P_2=\Pr[\text{1 node in } \mathcal A_2]=\pi\lambda_h((d_m+{\rm d}r_1)^2-d_m^2)\exp(-\pi\lambda_h((d_m+{\rm d}r_1)^2-d_m^2))$, $P_3=\Pr[\text{$M-K_n-m+1$ nodes in } \mathcal A_3]=\frac{(\pi\lambda_h (d_{M-K_n+1}^2-(d_m+{\rm d}r_1)^2))^{M-K_n-m+1}}{(M-K_n-m+1)!}\exp(-\pi\lambda_h (d_{M-K_n+1}^2-(d_m+{\rm d}r_1)^2))$ and $P_4 = \Pr[\text{1 node in } \mathcal A_4]=\pi\lambda_h((d_{M-K_n+1}+{\rm d}r_2)^2-d_{M-K_n+1}^2)\exp(-\pi\lambda_h((d_{M-K_n+1}+{\rm d}r_2)^2-d_{M-K_n+1}^2))$.
By \eqref{eqn:node-m-in-A-2}, we have
\small{\begin{align}
&f_{d_m,d_{M-K_n+1}}(x,y)=\lim_{{\rm d}r_1,{\rm d}r_2\to 0}\frac{\Pr[\text{node $m$ in } \mathcal A_2,\ \text{node $M-K_n+1$ in } \mathcal A_4]}{{\rm d}r_1{\rm d}r_2}\notag\\
&=\begin{cases}
\frac{4(\pi\lambda_h)^{M-K_n+1}d_m^{2m-1}d_{M-K_n+1}(d_{M-K_n+1}^2-d_m^2)^{M-K_n-m}\exp(-\pi\lambda_h d_{M-K_n+1}^2)}{(m-1)!(M-K_n-m)!} \hspace{5mm} d_m<d_{M-K_n+1}\\
0 \hspace{108mm} \text{otherwise}
\end{cases}\label{eqn:f_in_appendix}.
\end{align}}\normalsize
Substituting \eqref{eqn:PZF_cond_case1} and \eqref{eqn:g_in_appendix} into \eqref{eqn:PZF_STP_case1} and substituting \eqref{eqn:PZF_cond_case2} and \eqref{eqn:f_in_appendix} into \eqref{eqn:PZF_STP_case2}, we can obtain $\Pr\left[\text{SIR}_n^{\text{pzf}}>\tau|\ell_{0,n}=m\right]$ and $\Pr\left[\text{SIR}_n^{\text{pzf}}>\tau|\ell_{0,n}>M-K_n\right]$, respectively. Then, based on \eqref{eqn:PZF_STP_n_appendix}, we can obtain $q^{\text{pzf}}_{M,n}(K_n,T_n)$.
According to total probability theorem, we can show Theorem~\ref{Thm:STP_PZF_exact}.

\begin{center}
Appendix G: Proof of Lemma~\ref{Lem:upper_bound_PZF}
\end{center}
By the law of total probability, we have

\small{\begin{align}
q^{\text{pzf}}_{M,n}(K_n,T_n)=&\sum_{m=1}^{\infty}\Pr\left[\ell_{0,n}=m\right]\Pr\left[\text{SIR}_n^{\text{pzf}}>\tau|\ell_{0,n}=m\right]
\stackrel{(a)}{\leq}\sum_{m=1}^{M-K_n+L-1}\Pr\left[\ell_{0,n}=m\right]\Pr\left[\text{SIR}_n^{\text{pzf}}>\tau|\ell_{0,n}=m\right]\notag\\
&+\sum_{m=M-K_n+L}^{\infty}\Pr\left[\ell_{0,n}=m\right]\Pr\left[\text{SIR}_n^{\text{pzf}}>\tau|\ell_{0,n}=M-K_n+L\right],\notag\\
&\stackrel{(b)}{\leq}\sum_{m=1}^{M-K_n+L-1}T_n(1-T_n)^{m-1}R_{M,K_n,m}+(1-T_n)^{M-K_n+L-1}R_{M,K_n,M-K_n+L}\notag
\end{align}}\normalsize
where (a) is due to $\Pr\left[\text{SIR}_n^{\text{pzf}}>\tau|\ell_{0,n}=m\right]>\Pr\left[\text{SIR}_n^{\text{pzf}}>\tau|\ell_{0,n}=k\right]$ for any $k>m$, (b) is due to the fact that $R_{M,K_n,m}$ is an upper bound on $\Pr[\text{SIR}_n^{\text{pzf}}>\tau|\ell_{0,n}=m]$ obtained by adopting the lower bound on the incomplete gamma function, i.e., $\left(1-e^{-S_ab}\right)^a<\frac{\gamma(a,b)}{\Gamma(a)}$ \cite{alzer1997some}. 
$R_{M,K_n,m}$ can be calculated in a similar way to $q_{M,n}^{\text{mrc},u}(T_n)$ in Lemma~\ref{Lem:bounds_MRC}. We omit the details due to page limitation. Therefore, we complete the proof of Lemma~\ref{Lem:upper_bound_PZF}.

\vspace{-4mm}
\section*{Appendix H: Proof of Lemma~\ref{Lem:PZF_properties_UB}}

By denoting $P_n=1-T_n$, we have 
\small{\begin{align}
q^{\text{pzf},u}_{M,n}(K_n,1-P_n)&=\sum_{m=1}^{M-K_n+L-1}(1-P_n)P_n^{m-1}R_{M,K_n,m}+P_n^{M-K_n+L-1}R_{M,K_n,M-K_n+L}\notag\\
&=R_{M,K_n,1}-\sum_{m=1}^{M-K_n+L-1}P_n^{m}(R_{M,K_n,m}-R_{M,K_n,m+1}).\notag
\end{align}}\normalsize
Since $R_{M,K_n,m}-R_{M,K_n,m+1}>0$, $q^{\text{pzf},u}_{M,n}(K_n,1-P_n)$ is a decreasing function of $P_n$. In addition, since  $q^{\text{pzf},u}_{M,n}(K_n,1-P_n)$ is not affected by any $P_{n'}$, $n'\neq n$ and $\frac{{\rm d}^2 q^{\text{pzf},u}_{M,n}(K_n,1-P_n)}{{\rm d} P_n^2}<0$, $q^{\text{pzf},u}_M(\mathbf K,1-\mathbf P)$ is a concave function of $\mathbf P$. Therefore, $q^{\text{pzf},u}_M(\mathbf K,\mathbf T)$ is a concave increasing function of $\mathbf T$.


\vspace{-3mm}

\end{document}